\documentclass{article}

\usepackage{PRIMEarxiv}

\usepackage[utf8]{inputenc} % allow utf-8 input
\usepackage[T1]{fontenc}    % use 8-bit T1 fonts
\usepackage{hyperref}       % hyperlinks
\usepackage{url}            % simple URL typesetting
\usepackage{booktabs}       % professional-quality tables
\usepackage{nicefrac}       % compact symbols for 1/2, etc.
\usepackage{microtype}      % microtypography
\usepackage{lipsum}
\usepackage{fancyhdr}       % header
\usepackage{graphicx}       % graphics
\graphicspath{{media/}}     % organize your images and other figures under media/ folder

\usepackage{cite}
\usepackage{amsmath,amssymb,amsfonts}
\usepackage{algorithmic}
\usepackage{subcaption}
\usepackage{textcomp}
\usepackage{array,multirow}

\usepackage{xcolor}

\title{DICTION:	DynamIC robusT whIte bOx watermarkiNg scheme for deep neural networks
}

\author{
  Reda Bellafqira \\
 IMT Atlantique\\
  Inserm, UMR 1101 Latim\\
  29238 Brest Cedex, France \\
  \texttt{reda.bellafqira@imt-atlantique.fr} \\
   \And
  Gouenou Coatrieux \\
 IMT Atlantique\\
  Inserm, UMR 1101 Latim\\
  29238 Brest Cedex, France \\
  \texttt{gouenou.coatrieux@imt-atlantique.fr} \\
}

\begin{document}
\maketitle

\begin{abstract}
 Deep neural network (DNN) watermarking is a suitable method for protecting the ownership of deep learning (DL) models. It secretly embeds an identifier (watermark) within the model, which can be retrieved by the owner to prove ownership. In this paper, we first provide a unified framework for white box DNN watermarking schemes. It includes current state-of-the-art methods outlining their theoretical inter-connections. Next, we introduce DICTION, a new white-box Dynamic Robust watermarking scheme, we derived from this framework. Its main originality stands on a generative adversarial network (GAN) strategy where the watermark extraction function is a DNN trained as a GAN discriminator taking the target model to watermark as a GAN generator with a latent space as the input of the GAN trigger set. DICTION can be seen as a generalization of DeepSigns which, to the best of our knowledge, is the only other Dynamic white-box watermarking scheme from the literature. Experiments conducted on the same model test set as Deepsigns demonstrate that our scheme achieves much better performance. Especially, with DICTION, one can increase the watermark capacity while preserving the target model accuracy at best and simultaneously ensuring strong watermark robustness against a wide range of watermark removal and detection attacks. 

\end{abstract}

% keywords can be removed

\maketitle

\section{Introduction}
\label{sec:intro}
Deep Learning (DL) has enabled significant and rapid progress in signal analysis and applications such as speech recognition, natural language processing, diagnostic support, and so on. But, more and more, building a deep learning model is an expensive process that requires: (i) a lot of data; (ii) a lot of computational resources (like GPUs); (iii) the help of DL experts to carefully define the network topology and correctly set the training hyper-parameters (like learning rate, batch size, weights decay, etc.); (iv) the help of experts in the field of application, sometimes with very rare knowledge as in the medical domain, for instance. Because of this, the illegal copying and redistributing of such a model is a big financial loss for its creators.

Watermarking has been proposed to protect the ownership of deep learning models. It consists of inserting a message (a watermark) in a host document by imperceptibly modifying some of its characteristics. For example, the watermarking of an image \cite{bouslimi2016data} is based on the slight modification (or modulation) of its pixel values to encode the message. This message can then help verify the origin/destination of the document it protects in the context of copyright protection or fighting data leaks \cite{niyitegeka2018dynamic, aberna2023digital}. The Watermarking of DL models relies on the same principles, taking of course into account their specific features and operating constraints. A DL model differs from a multimedia document in many ways.

Designing an effective deep watermarking scheme needs to meet the following six key criteria: 
\begin{itemize}
\item \textbf{Fidelity} - which means that the performance of the model should be well preserved after watermarking;  
\item \textbf{Integrity} - which refers to the watermarking scheme's ability to produce minimal false alarms and ensure that the watermarked model can be uniquely identified using the appropriate watermarking key; 
\item \textbf{Robustness} - which means that the watermark cannot be easily removed when the model is attacked; 
\item \textbf{Security} - which guarantees that the watermark cannot be easily detected by malicious attackers; 
\item \textbf{Capacity} - which refers to the scheme's ability to effectively embed an amount of data (payload - size of the watermark) into the protected model, and; 
\item \textbf{Efficiency} - which is the need for the computation cost of watermark embedding and extraction to be kept reasonably low.
\end{itemize}
In practice, most schemes achieve a compromise between these criteria. For instance, the watermark should not degrade the performance of the model while being at the same time resistant to attacks aimed at detecting or removing it. 
 Among these attacks, one can cite the classic ones:
\begin{itemize}
    \item The pruning attack (PA), where model weights whose absolute values are below a threshold are set to zero. 
    
    \item The fine-tuning attack (FTA), which re-trains and updates the model without decreasing its accuracy.
    
    \item The overwriting attack (OWA) – where the attacker embeds his/her watermark to suppress the original one.
    
    \item The Wang and Kerschbaum attack (WKA) which checks the weight distribution of the watermarked model.  
    
    \item The property Inference Attack (PIA) trains a discriminator capable of differentiating watermarked from non-watermarked models, capturing watermarked weights, and determining whether the model under attack is no longer watermarked.  
\end{itemize}
Notice that removal attacks are efficient if the attacked model also has a high test accuracy and if its runtime and needed resources are small compared to retraining a model from scratch.

In the sequel, we come back to related works before introducing the contributions of this paper: a unified framework of DL watermarking; and DICTION a new and efficient scheme.

\subsection{Related work}
One can classify DNN watermarking methods into black-box (BBx) and white-box (WBx) schemes \cite{xue2021dnn, lukas2022sok, li2021survey, fkirin2022copyright, boenisch2021systematic, sun2023deep}. By definition, a black-box watermark can be extracted by querying only the model. The watermarked message is read by looking at the outputs of the watermarked model for some well-designed and crafted inputs referred to as the "trigger set" \cite{chen2019blackmarks, vybornova2022method, zhang2018protecting, adi2018turning, guo2018watermarking, le2020adversarial, namba2019robust, li2019prove, kapusta2020watermarking, lounici2022blindspot, kallas2022rose, qiao2023novel, kallas2023mixer, hua2023unambiguous}. These methods have a zero-bit capacity, as they can only detect the presence of a given watermark in a protected model. This is equivalent to extracting a message $b$ of one bit equal to '1' if the watermark is detected or '0' otherwise. 

On their side, white box (WBx) watermarking schemes \cite{uchida2017embedding, feng2020watermarking, li2020spread, tartaglione2021delving, chen2019deepmarks, wang2020watermarking, rouhani2019deepsigns, wang2020watermarking, fan2019rethinking} require the access to the whole model parameters to extract the watermark. These schemes have a multi-bit capacity. They allow the encoding of a message containing several bits of information. In other words, the message $b=\{m_i\}_{i=1 ..N}$, with $ m_i\in \{0, 1\}$ can be chosen by the user. WBx schemes can be differentiated depending on whether the watermark is read directly from the network weights; in that case, they are called "static watermarking" schemes; or, by examining the values of the intermediate activation maps for specific inputs of the model, they are called "dynamic watermarking" schemes \cite{rouhani2019deepsigns}. 

If we go deeper into the classification of WBx watermarking schemes, one can differentiate them according to whether they apply embedding modulations as with multimedia data, or whether they change the way the model is trained by modifying its training loss, for example. The first class involves modifying a set of secretly selected model parameters using a classic multimedia watermark modulation and, to preserve the accuracy of the model, fine-tuning the model;  updating non-watermarked parameters until the model recovers its accuracy.

For example, in \cite{song2017machine}, Song et al. embed the watermark in the least significant bits or signs of the model parameters. They developed a correlated value encoding method to maximize the correlation between the model parameters and a given secret. Feng et al. \cite{feng2020watermarking} randomly select some model's weights on which they apply an orthogonal transformation. The watermark is then embedded using a binarization method on the obtained coefficients and the inverse orthogonal transformation is applied. Fine-tuning is finally done on the non-watermarked weights.\cite{kuribayashi2020deepwatermark} works similarly and embeds the watermark into the discrete cosine transform (DCT) coefficients of secretly selected weights using the dither modulation and quantization index modulation (DM-QIM) method. Tartagoline \cite{tartaglione2021delving} proposed a different approach based on using a real-valued watermark instead of a binary string. The watermark is embedded into a secret subset of the model parameters using an affine transformation and replica-based approach. The weights carrying the watermark are not updated during training, allowing for easy extraction.

As stated, the second class of watermarking techniques involves modifying the loss function of the model by adding a regularization term to it. This term allows for a trade-off between the strength of the watermark and the overall performance of the model unlike the previous works. More clearly, by adjusting the regularization hyperparameter, one can control the balance between embedding a strong watermark and maintaining the model's accuracy. This flexibility ensures that the watermark is effectively embedded without sacrificing the model's predictive capabilities. Although we return to these schemes in detail to unify them in a single DL watermarking framework in Section \ref{sec:related_work}, we feel it important for us to give an overview of their technical evolution. 

Uchida et al. \cite{uchida2017embedding} proposed one of the first explicit watermarking schemes of this kind. They insert a binary string as a watermark in the model parameters by training or fine-tuning the model using a loss function composed of the model's original loss along with an embedding regularizer. An embedding parameter weights the regularizer to manage the trade-off between watermark robustness and preservation of model accuracy. However, this method is not robust to overwriting attacks. Li et al. \cite{li2020spread} extended this approach and proposed a DNN watermarking algorithm based on spread-transform dither modulation (ST-DM). The DNN is trained with a loss function explicitly designed with a regularizer to ensure the correlation of the DNN weights with a given spreading sequence.  \cite{li2022encryption} propose to insert the watermark into fused kernels instead of using them separately to resist parameter shuffling attacks. Because this strategy reduces the capacity, the authors introduce a capacity expansion mechanism using a MappingNet to map the fused kernels into a higher dimension space capable of hosting the watermark. However, it should be noted that the shuffling attack is not efficient, since it degrades the model performance. Wang and Kerschbaum  \cite{wang2019attacks} demonstrated that the aforementioned approaches based on Uchida et al. \cite{uchida2017embedding} scheme do not fulfill the requirement of watermark secrecy because they introduce easily detectable changes in the statistical distribution of the model parameters. To address this limitation, Wang and Kerschbaum \cite{wang2019riga} proposed a strategy to create undetectable watermarks in a white box setting using generative adversarial networks (GAN). In this approach, the watermarked model serves as the generator while the watermark detector, which acts as a discriminator, detects changes in the statistical distribution of the model parameters.

Unlike previous studies, different methods for verifying the ownership of deep neural networks based on passports have been proposed in \cite{fan2019rethinking, zhang2020passport}. That indeed increases the connection between network performance and correct passports by embedding passports in special normalization layers, resulting in enhancement of the watermarks resisting ambiguity attack. This attack consists of forging a new watermark on a given model without necessarily preventing the legitimate model owner from successfully verifying the watermark in a stolen model instance. However, it creates an ambiguity in which, for an external entity, such as a legal authority, it is not possible anymore to decide which party has watermarked the given model. Thus, the attack prevents the legitimate owner from successfully claiming the copyright of the intellectual property. This attack can be solved by either using an authorized time server or a decentralized consensus protocol with which an author authorizes its time-stamp to the verification community before publishing the DNN model \cite{li2021secure}. 

Rouhani et al. \cite{rouhani2019deepsigns} introduced DeepSigns. It is the first scheme that embeds a sequence of bits as a watermark into the probability density function (PDF) of the activation maps of multiple layers in a model, contrary to the previous schemes that work with the model's parameters. This is accomplished through a fine-tuning process that relies on a trigger dataset; a secretly selected subset of input samples; and the use of specialized regularizers to balance model accuracy and watermark detection. The fundamental idea behind DeepSigns is to cluster the activation maps of the model training set and embed the watermark within a secret subset of the clustered classes. To insert the message, the mean of each of these classes is computed on the trigger dataset and modified through fine-tuning such that its projection in a secret space defined by a matrix matches the watermark. The extraction procedure only requires knowledge of the trigger dataset, calculating the statistical mean of each class' activation maps, and their projection in the secret space to extract the watermark. Although DeepSigns achieves good model accuracy after watermarking, it suffers from poor insertion capacity and significant vulnerabilities to certain attacks such as FTA, PA, and OWA as the size of the watermark increases. 

Table \ref{tab:my_label} summarizes the above scheme taxonomy according to whether they are white and/or black box, whether they are static or dynamic, how/when the watermark is inserted, and the secret watermark key parameters. Notice also that these algorithms have been proposed considering centralized learning. Some algorithms have been proposed in the context of federated learning where the training is shared between a group of clients having their own data they don't want to share for privacy concerns. Such schemes are extensions of the above schemes. We invite the reader to refer to \cite{lansari2023federated}.

\begin{table*}
    \centering
    \begin{tabular}{|c|c|c|c|c|c|c|c|}
       \hline
       \multirow{2}{*}{Existing works}            & \multicolumn{2}{c|}{White box}  & Black box     & \multicolumn{2}{c|}{Stages of application} & \multirow{2}{*}{Secret key}  \\
       \cline{2-6}
                                                    & Dynamic & Static                &  Dynamic      & Training & Fine-tuning     &   \\
        \hline
        Uchida et al. \cite{uchida2017embedding} &        &   $\checkmark$      &             &     $\checkmark$         & $\checkmark$   & Matrix   \\
        \hline
        DeepSigns \cite{rouhani2019deepsigns} &    $\checkmark$     &        &   $\checkmark$             &             & $\checkmark$ & TriggerSet, Matrix \\
        \hline
        RIGA \cite{wang2019riga} &         &   $\checkmark$     &          &    $\checkmark$              &  $\checkmark$   & DNN \\
        \hline
                Li et al. \cite{li2020spread}  &         &   $\checkmark$       &          &             $\checkmark$   &  $\checkmark$   &  Matrix  \\
        \hline
                Tartaglione et al. \cite{tartaglione2021delving} &       &       $\checkmark$     &        &         &$\checkmark$  &  Matrix \\
        \hline
                Kuribayashi et al.  \cite{kuribayashi2020deepwatermark}  &         &     $\checkmark$   &           &           &  $\checkmark$ & Matrix  \\
        \hline
                Feng et al.  \cite{feng2020watermarking}                 &         &   $\checkmark$   &            &            &   $\checkmark$  & Matrix  \\
        \hline
             EncryptResist \cite{li2022encryption}  &         &  $\checkmark$      &          &    $\checkmark$       &   $\checkmark$            &  DNN, Matrix \\
        \hline
             Fan et al. \cite{fan2019rethinking}  &         &  $\checkmark$      &    $\checkmark$      &      $\checkmark$     &    $\checkmark$  & Passport layer and Matrix   \\
        \hline
                  DICTION (our work)  &    $\checkmark$     &        &          &     $\checkmark$     &    $\checkmark$    &  TriggerSet, DNN  \\
        \hline
    \end{tabular}
    \caption{
 Taxonomy of ML watermarking schemes, classified according to : whether they are white or black box; when the watermark is inserted (training, fine-tuning or after the model has been trained); and, the watermarking key parameters. }
    \label{tab:my_label}
\end{table*}

\subsection{Motivation and contribution}

In the literature, there is a lack of a formal definition of white box watermarking scheme, as well as a watermarking scheme that meets all the requirements mentioned above. To this end, we present a formal definition of a white box watermarking algorithm. And then, we propose a novel dynamic white-box algorithm DICTION that overcomes the limitations cited above. In particular, we propose to use a neural network following a generative adversarial-based strategy to map the activation maps to the watermark space rather than a static projection matrix as DeepSigns. Indeed, using a static projection matrix to ensure that the watermarked activation maps' clustering matches the watermark in the projected domain does not guarantee that this clustering respects that of the original model after DeepSigns has fine-tuned the model for watermark insertion. 

To embed a watermark $b$ in a target model, DICTION first considers the initial portion of the target model up to layer $"l"$ as a generative adversarial network (GAN) generator \cite{goodfellow2014generative} and then trains a projection neural network (like a GAN discriminator) concurrently with the generator model in such a way they map the trigger set activation maps of the watermarked model to the desired watermark $"b"$ and those of the non-watermarked model (i.e. the original one) to random watermarks. To detect the watermark, it is thus sufficient to take the trigger set samples as model input and to verify that the projection neural network reconstructs the watermark from the corresponding activation maps. Using a neural network for activation-maps-watermarking space mapping significantly expands the ability to embed a watermark while also increasing the watermark capacity. Additionally, our neural network for projection can be trained while embedding the watermark into a pre-trained model or not. DeepSigns only fine-tune the target model. Another originality of our proposal is that it makes use of a trigger set made up of samples issued from a latent space defined by its mean and standard deviation and that it can insert a watermark into the activation maps of a given layer or a combination of multiple layers. More clearly, the concatenation of these activation maps will be in the form of a tensor, where each channel is represented by an output from a hidden layer of the model. In such a strategy, the watermark will be propagated to all layers of the model and better preserve the model accuracy, avoiding possible misclassification of real images as for DeepSigns. It also simplifies the detection as one just has to generate samples of the latent space to constitute the trigger set. DICTION is thus independent of the selection of a set of real samples, e.g. images, as a trigger set. Moreover, very few parameters have to be stored (the mean and the standard deviation). In some cases, model accuracy is improved because our watermarking proposal is considered a regularization term. As we will see, our proposal is very resistant to all the attacks listed previously (PA, FTA, OA, WKA, and PIA). Outperforms the DeepSigns white-box dynamic watermarking scheme in terms of the robustness of the watermark and the precision of the watermarked model.

The rest of this paper is organized as follows. In Section \ref{sec:related_work}, we present the most recent white-box watermarking schemes, static or dynamic, illustrating their advantages and weaknesses by including them in a unified white-box watermarking framework; another contribution of this work. We then take advantage of this framework to introduce in Section \ref{sec:diction} our proposal: DICTION, a novel dynamic white box watermarking scheme that is a generalization of DeepSigns with much better performance. Section \ref{sec:experiments} provides experimental results and a comparison assessment of our scheme with state-of-the-art solutions. Section \ref{sec:conclusion} concludes this paper.

%\section{Related works and focus on DeepSigns}

\section{Unified WBx ML watermarking framework}

\label{sec:related_work}

In this section, we provide a formal definition of white-box neural network watermarking, a unified framework, by presenting the two most recent and efficient static schemes \cite{wang2019riga, uchida2017embedding, li2022encryption} and, to our knowledge, the one dynamic scheme \cite{rouhani2019deepsigns}. Before going into the details of these approaches, we first give a classical definition of the architecture of deep neural networks, which will serve as a reference in the following.

\subsection{Deep Neural Networks (DNNs)}

\begin{figure*}
     \centering
     \includegraphics[scale=.5]{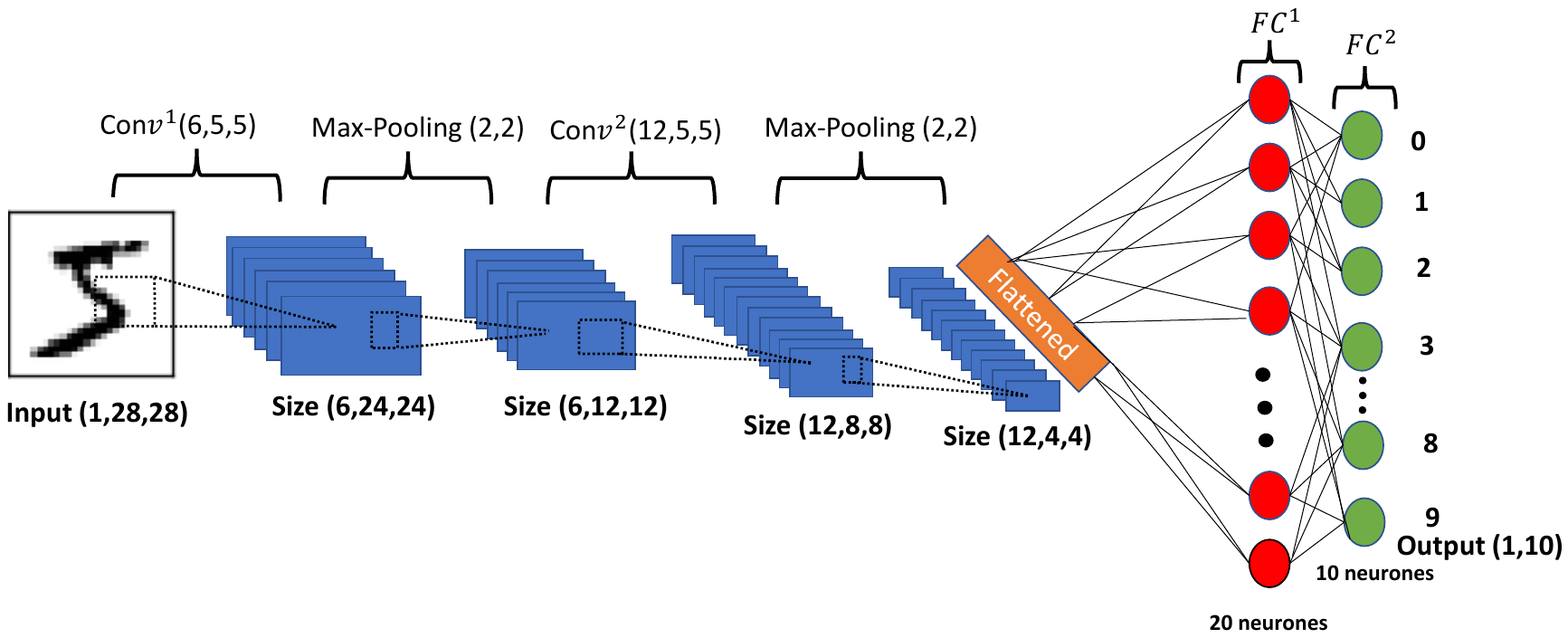}
     \caption{Example of a Convolutional Neural Network architecture for image classification (using the well-known image dataset MNIST \cite{lecun1998gradient}), with 2 convolution layers, 2 max-pooling layers, and a fully connected neural network of two layers.}
     \label{fig:cnn}
 \end{figure*}

 In this work, we decided to consider a classical DNN architecture for image classification due to the fact the vast majority of neural network watermarking schemes have been developed and experimented with in this context. As depicted in Fig. \ref{fig:cnn}, such a DNN model $M$ is a stack of layers that transforms an input layer, holding data $X$ to classify, into an output layer holding the label score $Y$. Different kinds of layer are commonly used:

\begin{enumerate}
    \item Fully connected (FC) layer  – It consists of a set of neurons such that each neuron of a layer $\text{FC}^{m}$ is connected to all those of the previous layer $\text{FC}^{m-1}$. More clearly, a  FC layer of $I_m$ neurons connected to a previous layer of $I_{m-1}$ neurons will receive as input a vector $a^{(m-1)}$ of size $I_{m-1}$ and will provide as output a vector $a^{(m)}$ with $I_m$ components that will be the input of the next layer. Such computation can be expressed as a mapping of the form:
    \begin{equation}
       a^{(m)} = g_m(W^{(m)} a^{(m-1)} +b^{(m)}) 
    \end{equation}
     where: $W^{(m)}$ is the matrix of neuron weights of shape $I_m\times I_{m-1}$; $b\in \mathbb{R}^{I_m}$ is the neurons' bias vector; and, $g_m()$ is a nonlinear activation function applied component-wise (e.g, Relu, Sigmoid, Tanh ...). 

    \item  Convolutional layer $( \text{Conv}(O_m, P_m, Q_m)$): – As illustrated in Fig. \ref{fig:cnn}, such a layer receives as input an "image" $A^{(m-1)}$ of $O_{m-1}$ channels and computes as output a new "image" $A^{(m)}$ composed of $O_m$ channels. The output at each channel is usually named activation maps and is computed as:
     \begin{equation}
         A_0^{(m)} = g_m(\sum_k W_{ok}^{(m)}*A_k^{(m-1)} + b_0^{(m)})
     \end{equation}
     where: $"*"$ denotes the 2D convolution operation; $W_{ok}^{(m)}$ is the filter weight matrix of shape $P_m \times Q_m$ along with a bias $b_0^{(m)} \in \mathbb{R}$. The matrix $W_{ok}^{(m)}$ parameterizes a spatial filter that is automatically tuned during training to detect or enhance some features from $A^{(m-1)}$.
      \item Pooling $(\text{PL})$ layers – It is implemented in a DNN as a spatial dimensionality reduction operation. The most common pooling layers are max pooling and average pooling. Roughly, Max-Pooling($n$,$n$) is a patch kernel operator of size $n \times n$ applied to $A^{(m)}$ that outputs the maximum of the values found in each $n \times n$ patches of $A^{(m)}$. For average pooling, one takes the average of the values in a patch. The right choice of this function can make the model more robust to distortions in the input pattern. Anyway, a PL layer reduces the number of trainable parameters for the next layers. 

    \item Flatten Layer – it corresponds to a special reshaping operation that is required to move from bidimensional (or multidimensional) layers to one-dimensional fully connected layers.
\end{enumerate}

Architecting a DNN entails developing a suitable sequence of convolutional, pooling, and fully connected layers, as well as their hyperparameters (e.g., learning rate, momentum, weight decay, etc.). Typical architectures, as shown in Fig. \ref{fig:cnn}, introduce a pooling layer after one convolutional layer, defining a convolutional block that is repeated until the activation maps size is small enough to feed fully connected layers. A special reshaping operation known as "a flattened layer" is required to transition from bidimensional (or multidimensional) layers to one-dimensional fully connected layers.
 
 The training of DNN is based on two iterative phases: feed-forward and back-propagation. Before the first feed-forward phase begins, all model parameters are initialized with random values, for example. The training data are then fed into the DNN according to a batch size which is the number of samples used to perform a single forward and backward pass. The difference between the original label $(Y_{\text{Train}})$ and the ones computed by the DNN is then calculated using an objective function (also known as the loss function $(E(.))$) (e.g. cross entropy for classification, Mean Square Error for regression). This error is then used in the back-propagation phase to update all model parameters using techniques such as gradient descent. 

 Once the DNN model is trained, that is, once the model parameters are known (e.g., model accuracy does not evolve anymore), it can be used to classify new data (inference phase). This classification process simply consists of applying the feed-forward phase with new data as input, considering that the DNN output will give the class of the input data.

\subsection{White-Box watermarking}
\label{subsec: whiteWat}

As stated above, there are two classes of white box watermarking schemes: the static class and the dynamic class. The former groups the methods that embed the watermark by directly modifying the weights of a trained model \cite{feng2020watermarking, kuribayashi2021white} using a classical watermarking modulation as for images. In the second class \cite{uchida2017embedding, rouhani2019deepsigns, li2020spread, wang2019riga, li2022encryption}, the watermark is inserted during the training of the model for the original task it is designed for. By doing so, the idea is that the watermark insertion process does not reduce the performance of the host model. From our point of view, these methods can be integrated within a unified white-box watermarking framework that depicts any of these schemes accordingly in the two following main steps:

\begin{enumerate}
    \item 	At first, considering a target model  $M$, a secretly parameterized features extraction function $\text{Ext}(M,K_{\text{ext}})$ is applied using a secret key $K_{\text{ext}}$. As we will see, such features can be a subset of the model weights. In that case, $K_{\text{ext}}$ simply indicates the indices of the model weights to select. It can also be some model activation maps for some secretly selected input data, issued from a trigger set. These features are then used for the insertion / extraction of the watermark.
    \item 	The embedding of a watermark message $b$ consists of regularizing $M$ with a specific watermarking regularization term $E_{\text{wat}}$ added in the model loss so that the projection function $\text{Proj}(.,K_{\text{Proj}})$ applied to the selected features encodes $b$ in a given watermark space; space which depends on the secret key $K_{\text{Proj}}$. The idea is that after training we have the following: 
           \begin{equation}
              \text{Proj(Ext}(M^{\text{wat}}, K_{\text{ext}}), K_{\text{Proj}}) = b 
           \end{equation}
    where $M^{\text{wat}}$ is the watermarked version  model of the target model $M$.

The watermarking regularization term $E_{\text{wat}}$ depends on a distance measure $d$ defined in the watermark space, like the hamming distance, Hinge Loss or cross-entropy in the case of a binary watermark that is a binary string of length $l$, i.e., $b \in \{0,1\}^l$). $E_{\text{wat}}$ is thus defined as: 

 \begin{equation}\label{eq:loss_wat}
E_{\text{wat}} = d(\text{Proj(Ext}(M^{\text{wat}}, K_{\text{ext}}), K_{\text{Proj}}), b)
\end{equation}
To preserve the accuracy of the target model, the watermarked model $M^{\text{wat}}$  is in general derived from $M$ through a fine-tuning operation parametrized with the following loss function:
\begin{equation}
   E = {E}_{0}(X_{\text{Train}}, Y_{\text{Train}}) + \lambda E_{wat} 
\label{eq:loss_global_wat}
\end{equation}
 
where: $E_{0}(X_{\text{Train}}, Y_{\text{Train}})$ represents the original loss function of the network and $\lambda$ is a parameter for adjusting the trade-off between the original loss term and the watermarking regularization term. Herein, $E_0$ is important to ensure good behavior about the classification task while $E_{\text{wat}}$ ensures the correct insertion of the watermark.
\end{enumerate}

The watermark retrieval is by next pretty simple. It consists in using both the features extraction function $\text{Ext}(.,K_{\text{ext}})$ and the projection function $\text{Proj}(.,K_{\text{Proj}})$ as follow : 
\begin{equation}
    b^{\text{ext}} = \text{Proj(Ext}(M^{\text{wat}},K_{\text{ext}}),K_{\text{Proj}})
    \label{eq:wat_extract}
\end{equation}
where $b^{\text{ext}}$ is the extracted watermark or message from $M^{\text{wat}}$.

In the following sub-sections, we depict the algorithms of Uchida et al. \cite{uchida2017embedding}, RIGA \cite{wang2019riga}, ResEncrypt \cite{li2022encryption}  and Deepsigns \cite{rouhani2019deepsigns} accordingly to our framework. 

\subsubsection{Uchida et al. algorithm \cite{uchida2017embedding}}
In the scheme of Uchida et al. \cite{uchida2017embedding}, the feature extraction function $\text{Ext}(., K_{\text{ext}})$ corresponds to the mean value of some secretly selected filter weights. More clearly, based on $K_{\text{ext}}$, a convolutional layer is selected. Let $(s,s)$, $d,$, and $n$ represent the kernel size of the layer filters, the number of channels of the layer input, and the number of filters, respectively. Let also note the tensor $\mathbf{W} \in \mathbb{R}^{s\times s\times d\times n}$ with all the weights of these filter layers. The feature extraction function of the Uchida et  al.'s scheme then includes the following steps :
\begin{enumerate}
    \item calculate the mean values of filters’ coefficients at the same position $\overline{W}_{ijk} = \frac{1}{n} \sum_{h=1}^{n}W_{ijkh}$, getting $\overline{\mathbf{W}} \in \mathbb{R}^{s\times s\times d}$. 
   \item flatten $\overline{\mathbf{W}}$ to produce a vector $\mathbf{w} \in \mathbb{R}^{v}$ with  $v=s\times s\times d$.
\end{enumerate}

The Uchida et al. projection function $\text{Proj}(.,K_{\text{Proj}})$ has been designed to insert a $l$ bit long watermark $b \in \{0, 1\}^l$ and is defined as: 
\begin{equation}
\label{eq:uchida_proj}
    \text{Proj}(\mathbf{w}, K_{\text{Proj}}) = \sigma(\mathbf{w}A) \in \{0, 1\}^l
\end{equation}
where $K_{\text{Proj}} = A$ is a secret random matrix of size $(|w|,l)$ and $\sigma (.)$ is the Sigmoid function: 
\begin{equation}\label{sigmoid}
\sigma(x)=\frac{1}{1+e^{-x}}
\end{equation}
As distance $d$ for the watermarking regularizer $E_{\text{wat}}$, Uchida et al. \cite{uchida2017embedding} exploits the binary cross entropy: 
\begin{equation}\label{regulization}
	d(b,y) = -\sum_{j=1}^{l}(b_{j}\log(y_{j})+(1-b_{j})\log(1-y_{j}))
\end{equation}
Based on $d$, $\text{Proj}(., K_{\text{Proj}})$ and $\text{Ext}(.,  K_{\text{ext}})$, one can then compute the loss $E_{\text{wat}}$ as given in \eqref{eq:loss_global_wat} to watermark a target model $M$.

Herein, it is interesting to see that the projection function $\text{Proj}(.)$ \eqref{eq:uchida_proj} corresponds to a simple perceptron with Sigmoid as an activation function and a parameter set $\theta$ that includes: null biases and weights $|W|$ (with $|.|$ the cardinality operator). So, the embedding process of Uchida et al.'s scheme is equivalent to train a perceptron $\text{Proj}_{\theta}(.)$ on the database  $(A_i, b_i)_{i=1...l}$, where the $i^{\text{th}}$ column of $A$, $A_i$, corresponds to a training sample with as label $b_i$ that is the $i^{\text{th}}$ bit of $b$, with a batch size of size $l$, that is the number of columns of $A$ or equivalently the size of watermark. To sum up, $\text{Proj}_{\theta}(.)$ is trained so that the projection of $A$ on $\theta$ gives $b$, a watermark of  $l$ bits.

From there, one can reformulate and generalize the insertion process of Uchida et al. Let us consider a target model $M$, a message $b$, the secret $K_{\text{Proj}}=A$ (the secret matrix), and a single perceptron $\text{Proj}_{\theta}(.)$ the parameters of which are initialized with the values of the secretly extracted features (i.e. $\theta=\text{Ext}(M, K_{\text{ext}})$. The insertion of $b$ consists in finetuning the target model $M$ into $M^{\text{wat}}$ using the following global loss: 
\begin{multline}
   E = {E}_{0}(X_{\text{Train}}, Y_{\text{Train}}) \\ + \lambda (d(\text{Proj}_{\theta}(A), b) \\ + ||\theta - \text{Ext}(M^{\text{wat}}, K_{\text{ext}}) ||) 
\end{multline}

To clarify, $d(\text{Proj}_{\theta}(A), b)$ allows $\text{Proj}_{\theta}(.)$ to find its parameters $\theta$ that minimize the distance between $\text{proj}_{\theta}(A)$ and $b$, while the term $||\theta - \text{Ext}(M^{wat}, K_{ext}) ||$ allows minimizing the Euclidean distance between $\theta$ and the secretly selected weights $w^{\text{wat}}$ extracted from the watermarked model $M^{\text{wat}}$ using the feature extraction function $\text{Ext}(M^{\text{wat}}, K_{\text{ext}})$. $E_0$ is present to preserve the model's accuracy. The interest of our reformulation is that it allows the parameterization of $M^{\text{wat}}$ and $\text{Proj}_{\theta}(.)$ training in a different fashion (for example, we can use different learning rates and optimizers in the learning of $\theta$ and $\mathbf{w}^{wat}$). We experimentally verified that this statement obtained the same performance as the original scheme of Uchida et al. \cite{uchida2017embedding}. It is important to notice that both models $M^{\text{wat}}$ and $\text{Proj}_{\theta}(.)$ are trained simultaneously for fast convergence, and that only $M^{\text{wat}}$ is given to the recipient.

The extraction of the watermark can also be expressed in two ways. The first involves hard thresholding the result of the $\text{Proj}(.,.)$ function applied to $w^{\text{wat}}=\text{Ext}(M^{\text{wat}}, K_{\text{ext}})$: 
\begin{equation}
	b^{\text{ext}}_i =\begin{cases}
		1 &  \text{Proj}(w^{\text{wat}}, K_{\text{Proj}})_i \geq 0.5\\
		0 & \text{otherwise}.
	\end{cases}
	\label{eq:Uchida_ext_1}
\end{equation}
where $\text{Proj}(w^{\text{wat}}, K_{\text{Proj}})_i$ is the $i^{\text{th}}$ component of \\
$\text{Proj(Ext}(M^{\text{wat}}, K_{\text{ext}}), K_{\text{Proj}})$ and $b^{\text{ext}}_i$ is the $i^{\text{th}}$ bit of the  watermark extracted from $M^{\text{wat}}$.

The second expression relies on hard thresholding the output of the perceptron  $\text{Proj}_{\theta}(.)$ with as parameters $\theta = \text{Ext}(M^{\text{wat}}, K_{\text{ext}})$:
\begin{equation}
	b^{\text{ext}}_i =\begin{cases}
		1 &  \text{Proj}_{\text{Ext(M}^{\text{wat}}, K_{\text{ext}})}(A_i) \geq 0.5\\
		0 & \text{otherwise}.
	\end{cases}
	\label{eq:Uchida_ext_2}
\end{equation}
where $A_i$ is the $i^{th}$ column of the secret matrix $A=K_{\text{Proj}}$.

\paragraph*{Advantages and drawbacks of this approach:}

Uchida et al.’s scheme has been shown to be robust to fine-tuning and pruning attacks. It also provides a quite large watermark capacity. It is however not robust to watermark detection attacks such as the Wang and Kerschbaum attack (WKA) and the Property Inference Attack  (PIA). This is because it modifies the weight distribution, in particular that of the watermarked layer, in a way that can be easily detected. The watermark can thus be erased without knowing $K_{\text{ext}}$  (see \cite{TWang} for more details). It is also not resistant to the overwriting attack. An adversary just has to embed his/her watermark in $M^{\text{wat}}$ to remove the watermark. Indeed, even if the attacker uses a secret key matrix $A_{\text{attack}}$ of the same or greater dimension than $A$ to embed his message $b_{\text{attack}}$, $A_{\text{attack}} w^{\text{wat\_att}}$ being a simple linear projection, it has a low linear mapping capacity and a randomly generated $A_{\text{attack}}$ is likely to be very similar to the original matrix $A$ for some of the watermark bits; bits that an adversary will overwrite.

\subsubsection{Wang et al. scheme (RIGA) \cite{wang2019riga}} 

This scheme, named RIGA for Robust whIte-box GAn watermarking \cite{wang2019riga},  is static and was proposed to overcome the weaknesses of the previous approach \cite{uchida2017embedding} against WKA, PIA and OWA attacks. 

The main idea of RIGA is to use as a projection function a DNN: $\text{Proj}_{\theta}^{DNN}$ of parameters $\theta$, instead of the Uchida et al. perceptron  $\text{Proj}_{\theta}$,  and to associate it with another neural network $F_{\text{det}}$ which is used to preserve the distance between the distributions of the watermarked and the nonwatermarked weights to be robust to the WKA.  

Regarding the RIGA feature extraction function, it is close to the one of Uchida et al. \cite{uchida2017embedding} or any white box algorithm (see section \ref{subsec: whiteWat}). It works as follows: 
 \begin{itemize}
     \item As in \cite{uchida2017embedding}, it calculates the mean values $\overline{W}_{ijk} = \frac{1}{n} \sum_{h=1}^{n}W_{ijkh}$, i.e. the average value of the filters’ coefficients at the same position, getting $\overline{\mathbf{W}} \in \mathbb{R}^{s\times s\times d}$ and flattening $\overline{\mathbf{W}}$ into the vector  $\mathbf{w} \in \mathbb{R}^{v}$ with  $v=s\times s\times d$.
     \item From $\mathbf{w}$, RIGA secretly extracts as features a subset of elements from $w$ such as: $$\mathbf{w}_s=\text{Ext}(M,K_{\text{ext}})=\mathbf{w}K_{\text{ext}}$$
     where, $K_{ext}$ is the secret extraction key defined as a selection matrix of size $|\mathbf{w}_s |\times|\mathbf{w}|$  with a $1$ at position $(i,1)$, $(i+1,2)$ and so forth, and $0$ otherwise, where $i$ is the start index of a layer.      
 \end{itemize}
RIGA scheme then trains $\text{proj}_{\theta}^{\text{DNN}}$ with the dataset $((\mathbf{w}_s^{\text{wat}},b), (\mathbf{w}_s,b_r))$, where:  $\mathbf{w}_s^{\text{wat}}$ and $\mathbf{w}_s$ are the extracted features from the watermarked model $M^{\text{wat}}$ and the original model $M$, respectively; and, $b$ and $b_r$ are the owner watermark and a random watermark, respectively. Their idea behind using such a dataset is to ensure that $\text{proj}_{\theta}^{\text{DNN}}$ provides $b$ only for $\mathbf{w}_s^{\text{wat}}$ or a random watermark for any other models, otherwise. Indeed, training $\text{proj}_{\theta}^{\text{DNN}}$ using only $(\mathbf{w}_s^{\text{wat}}, b)$ may lead to a trivial function that ignores the input, that is to say, $\text{proj}_{\theta}^{\text{DNN}}$ will map any of its input to $b$. This is the case for the Uchida et al.'s scheme \cite{uchida2017embedding} which trains its perceptron with the training set $(A_i, b_i)_{i=1...l}$, and which is not robust to overwriting. It is important to notice that with $\text{proj}_{\theta}^{\text{DNN}}$, one can insert more structured watermarks than that of Uchida et al. which is a sequence of bits. In that case, one will have to replace the cross-entropy distance of the perceptron of the Uchida et al.’s scheme by the Mean Squared Error (MSE) in the case $b$ is an image. As defined, the projection function $\text{proj}_{\theta}^{\text{DNN}}$ has as an input layer size equals to $|\mathbf{w}_s |$ and an output layer of the size equal to that of the watermark. RIGA ensures that $\text{proj}_{\theta}^{\text{DNN}}(\text{Ext}(M^{\text{wat}},K_{\text{ext}}))=b$ with $K_{\text{ext}}$ secret extraction key. Expressed in our general framework, the watermark extraction function is such as (see (\ref{eq:wat_extract})):
\begin{multline}
    b^{\text{ext}} = \text{Proj}(\text{Ext}(M^{\text{wat}},K_{\text{ext}}),K_{\text{Proj}})= \\ \text{proj}_{\theta}^{\text{DNN}}(\text{Ext}(M^{\text{wat} },K_{\text{ext}}))
    \label{eq:wat_extract_riga}
\end{multline}
where, the secret projection key $K_{\text{proj}}$ corresponds to the set of parameters of $\text{Proj}_{\theta}^{\text{DNN}}$,i.e., $K_{\text{Proj}}=\theta$. Let us recall that for the Uchida et al.'s scheme \cite{uchida2017embedding}  $K_{\text{Proj}}$ is the projection matrix $A$ (see \eqref{eq:uchida_proj}). 
To go further, we can start to refine the RIGA watermark regularizer term as:
\begin{multline}
E'_{wat} = d(\text{Proj}_{\theta}(\text{Ext}(M^{\text{wat}}, K_{\text{ext}})), b) \\ + d(\text{Proj}_{\theta}(\text{Ext}(M, K_{\text{ext}})), b_r)    
\end{multline}
where $d$ is the distance between the embedded watermark on the targeted one $b$. As stated, if $b$ is an image, $d$ will correspond to MSE. 

This preliminary RIGA watermark regularizer $E'_{\text{wat}}$ is completed with a second term to counteract the WKA and PIA attacks. As stated, RIGA includes another DNN model $F_{\text{det}}$ whose role is to keep the distribution of the watermarked weights as close as possible to that of the non-watermarked weights. To do so, $F_{\text{det}}$ is trained with the following dataset $((w^{\text{wat}},1), (w,0))$. In that way, $F_{\text{det}}$ is a binary classifier that discriminates between watermarked and nonwatermarked weights. The loss function used to train $F_{\text{det}}$ is the binary cross entropy $E_{\text{det}}$ given as: 
 \begin{equation}
     E_{det} = log(F_{\text{det}}(w_s^{\text{wat}})) + log(1-F_{\text{det}}(w_s^{\text{ori}}))
 \end{equation}
where, $w_s^{\text{wat}}$ are the extracted watermarked features and $w_s^{\text{ori}}$ are the corresponding features extracted from the target model $M$.

To sum up, RIGA watermarks a target model $M$ into $M^{wat}$ through a fine-tuning operation under the global loss defined as:
\begin{multline}
E =  {E}_{0}(X_{\text{Train}}, Y_{\text{Train}})+\lambda E_{\text{wat}} \\
= {E}_{0}(X_{\text{Train}}, Y_{\text{Train}})+\lambda_1 E'_{\text{wat}} + \lambda_2 E_{\text{det}} \\
=  {E}_{0}(X_{\text{Train}}, Y_{\text{Train}})+\lambda_1 E'_{wat} \\-\lambda_2 \log (F_{det}(w^{wat}))
\label{eq:riga_emb}
\end{multline}
where, $\lambda_1$ and $\lambda_2$ are two adjusment parameters. Herein, $E_0$ serves the primary task of the model, $E_{\text{wat}}$ ensures the watermark embedding in $w^{\text{wat}}$ from $M^{\text{wat}}$ while $F_{\text{det}}$ is used to preserve the weight distribution of $M^{\text{wat}}$. Notice that in \eqref{eq:riga_emb}, $F_{\text{det}}$ is trained under $E_{det}$  after each training batch of $M^{\text{wat}}$ under $E$. Adding the term $-\lambda_2 \log (F_{det}(w^{\text{wat}})$ to $E$, makes the learning of the model $M^{\text{wat}}$ generates weights that maximize $F_{\text{det}} (w^{\text{wat}})$ or, more clearly, weights of $M^{\text{wat}}$ that are close to those of $M$. This is equivalent to a GAN-based strategy, where $M^{\text{wat}}$ is the generator and  $F_{\text{det}}$ the discriminator that classifies weights given in input as $'0'$, non-watermarked, or $'1'$, watermarked. Thus $F_{det}$ is trained to learn to separate the watermarked weights from the nonwatermarked ones. On the other hand, $M^{\text{wat}}$ is thus trained not only to achieve its classification and embedding task but also to fool $F_{\text{det}}$. Also notice that to ensure the correctness of $F_{\text{det}}$, the authors of RIGA recommend clamping its parameters and feeding it with the sorted weights of $M^{wat}$.

\paragraph*{Advantages and drawbacks of this approach :}  
RIGA has been shown resistant to fine-tuning, pruning, and overwriting attacks. It is also robust to the watermark detection attacks: PIA and WKA, due to the use of $F_{\text{det}}$. The security of RIGA depends on the knowledge of the parameters of $\text{Proj}_{\theta}$ that plays the role of the secret key ($K_{\text{Proj}}=\text{Proj}_{\theta}$). Thus, the more complex $\text{Proj}_{\theta}$; with the addition of more layers; the more difficult it is for an attacker to guess the neural network $\text{Proj}_{\theta}$. Moreover, due to the strong fitting ability of neural networks, $\text{Proj}_{\theta}$ can be mapped to a wide range of data types of watermark messages $b$, e.g., a binary string or even a 3-channel image. For different types of $b$, one will have however to choose the appropriate distance $d$ measured accordingly.

 One of the first drawbacks of RIGA is that it relies on training three models $M^{\text{wat}}$, $\text{Proj}_{\theta}$, and $F_{\text{det}}$. 
 Such training requires finding the right hyperparameters to make them converge at the same time; a task that is sometimes hard to implement in the case of very deep models. Another non-negligible complexity of RIGA is keeping the distribution of watermarked weights close to that of non-watermarked weights. This is done through the term $-\lambda_2 \log F_{\text{det}}(w^{\text{wat}})$ in  $E$ (see (\ref{eq:riga_emb})). This means minimizing $||w-w^{wat}||$ which is somewhat contradictory with the objective of RIGA's  $E_{\text{wat}}$ term that wants $\text{Proj}_{\theta}(w^{\text{wat}})==b$ and $\text{Proj}_{\theta}(\text{w})==b_r$. This is why RIGA watermarking needs a large number of epochs to converge.  
 Additionally, RIGA needs the parameters of many non-watermarked models to train  $F_{det}$. In the experiments carried out in \cite{wang2019riga}, $500$ non-watermarked models were used. This is extremely costly from the computation complexity point of view. To give an idea, considering the ImageNet-1k dataset, it takes around 14 days to complete 90 epochs for training ResNet-50 on an NVIDIA M40 GPU  \cite{you2017100}.

\subsubsection{Li et al. scheme (ResEncrypt)  \cite{li2022encryption}}

Proposed by Li et al. in 2022 \cite{li2022encryption}, ResEncrypt is a static white box watermarking scheme. Contrary to Uchida and RIGA schemes where the watermark is embedded in the means of filter parameters, it inserts into the means of the filter kernels to resist the parameter shuffling attack. However, such a strategy reduces the insertion capacity as kernel averaging shrinks the length of the sequence that can be used to host the watermark. To address this issue, Li et al.  \cite{li2022encryption} propose to use a DNN model, called MappingNet, to map the means or fused kernels into a space of higher dimension to increase the watermark capacity. MappingNet and the target model are jointly trained to conduct the final watermark embedding.

The feature extraction function of ResEncrypt \( \text{Ext}(., K_{\text{ext}}) \) works as follows. First, based on \( K_{\text{ext}} \), a convolutional layer is secretly selected. Let \( (s,s) \), \( d \), and \( n \) represent the kernel size of the layer filters, the number of channels of the layer input, and the number of filters, respectively. The tensor \( W \in \mathbb{R}^{s\times s\times d\times n} \) contains all the weights of these filter layers. Then, features are extracted according to the two following steps:
\begin{enumerate}
    \item Calculate the mean values of kernels’ coefficients at the same position \( \overline{W}_{ij} = \frac{1}{n\times d} \sum_{h=1}^{n} \sum_{k=1}^{d} W_{ijkh} \), getting \( \overline{\mathbf{W}} \in \mathbb{R}^{s\times s} \). 
    \item  Use a MappingNet model \( F_{map}(.) \) to map the fused kernel into a higher dimension space. \( F_{map}(.) \) is essentially a Multilayer Perceptron with \( \overline{\mathbf{W}} \) as input and \( \mathbf{W}^{map} \) as output, i.e.,
\[
    \mathbf{W}^{map} = F_{map}(\overline{\mathbf{W}}).
\]
\end{enumerate}

The projection function \( \text{Proj}(., K_{\text{Proj}}) \) of ResEncrypt is similar to the one used by Uchida et al. \cite{uchida2017embedding}. Considering a watermark \( b \in \{0, 1\}^l \) of length $l$ bits, this function is defined as:
\[
    \text{Proj}(\mathbf{W}^{map}, K_{\text{Proj}}) = \sigma(\mathbf{W}^{map} \times  A) \in \{0, 1\}^l
\]
where, \( K_{\text{Proj}} = A \) is a secret random matrix of size \( (|\mathbf{W}^{map}|,l) \) and \( \sigma(.) \) is the Sigmoid function. 

Binary cross entropy is used as a distance in the watermarking regularizer $E_{\text{wat}}$ of ResEncrypt. Based on \( d \), \( \text{Proj}(., K_{\text{Proj}}) \), and \( \text{Ext}(.,  K_{\text{ext}}) \), one can then compute the loss \( E_{\text{wat}} \) to watermark a target model \( M \), that is to say: 

\begin{multline}
     E_{\text{wat}} =  d( \text{Proj}(\mathbf{W}^{map}, K_{\text{Proj}}), b) + \\ d( \text{Proj}(\mathbf{W}^{\text{map}}_{\text{orig}}, K_{\text{Proj}}), b_r)    
\end{multline}

where $\mathbf{W}^{\text{map}}_{\text{orig}}$ is the expansion  by $F_{\text{map}}$ of the weights of the original model and $b_r$ is a random watermark. As in RIGA, the second term of $E_{\text{wat}}$ is added so that $F_{\text{map}}$ only learns to expand the target model weights to the watermark $b$ and the weights of the original model to $b_r$.

\paragraph*{Advantages and drawbacks of this approach :}
ResEncrypt is a DNN watermarking scheme that aims at resisting to the parameter shuffling attack by taking advantage of the fusion of all the kernels from the same layer to produce a host feature sequence. As exposed, this feature extraction is similar to that of Uchida et al. \cite{uchida2017embedding}, with the difference being that the mean is calculated at the level of the kernels instead of the filters. This averaging makes the watermark more resistant to kernel shuffling attacks. However, the shuffle attack Li et al. they considered in their work is not effective as it degrades the model's performance when it succeeds in removing the watermark. Moreover, being based on the same strategy as Uchida et al. \cite{uchida2017embedding}, ResEncrypt suffers from the same problems: the projection function, being a simple linear projection, has a low linear mapping capacity, and a randomly generated \( A_{\text{attack}} \) is likely to be very similar to the original matrix \( A \) for some of the watermark bits, bits an adversary will overwrite. A MappinNet model is used for the sake of capacity expansion and does not help to solving this problem.

\subsubsection{Rouhani et al. scheme (DeepSigns)  \cite{rouhani2019deepsigns}}
Unlike previous works, which directly embed the watermark into the static content of the target DNN model, i.e., its weights, DeepSigns embeds the watermark into the probability density functions (pdf) of the activation maps of different layers. The watermarking process is thus both data- and model-dependent, meaning that the watermark is embedded into the behavior of the DNN, its dynamic content, that can only be triggered by passing specific input data to the model. As previously, let us consider a target model $M$ devoted to a classification problem with $Y$ classes. In the sequel, we explain how DeepSigns works and unite with our general WBx framework defining its: feature extraction function $\text{Ext}(M,K_{\text{ext}})$, projection function $\text{Proj}(.,K_{\text{proj}})$ and its watermarking regularization term $E_{\text{wat}}$. 

DeepSigns watermarks the layers of $M$ independently. In a layer  $l$, it embeds a binary random watermark $b=\{b_i\}$, $b_i\in \{0,1\}$, by modulating some features of the layer outputs considering a given trigger set. These features are extracted as follows:

\begin{enumerate}
    \item Fed the model with all training data ($X_{\text{train}}, Y_{\text{train}}$) to get, for a given layer $l$ the corresponding activation maps:   $\{f^l(x, w)\}_{x \in X_{\text{train}}}$, where $f^l(x, w)$ belongs to $\mathbb{R}^m$, $m$ being the dimension of the activation maps (e.g. the number of neurons in the case of a $FC$ layer). 

   \item Define a Gaussian Mixture Model (GMM), of $S$ Gaussians, fitted the activation maps from step 1. Each sample $x$ of the training set therefore belongs to a GMM class $y_{gmm}$ and also to a classification class $y_{train}$, that is the label of $x$. The authors of DeepSigns set $S$ equal to the number of classes of the classification task of $M$.
    
    \item DeepSigns then secretly selects $s$ Gaussians from the mixture; Gaussian identified by their indices $T$ (e.g., if $S=10$, $T=\{1,5, 7\}$ then $s=3$); and computes their respective mean values $\{\mu_l^i\}_{i \in T}$, where $\mu_l^i \in \mathbb{R}^m$. These mean values will be used for the insertion of the watermark
    .
    \item  As the next step, DeepSigns secretly selects a subset of the input training data to build the trigger set of the layer $l$: $(X_{\text{key}}, Y_{\text{key}})$. These samples belong to the secretly selected Gaussian classes $(X_{\text{key}}, Y_{\text{key}})$ and the training labels of which also belong to $T$, more clearly: 
    \begin{multline}
        (X_{\text{key}}, Y_{\text{key}}) = \{(x, y_{\text{train}})\in \\ (X_{\text{train}},Y_{\text{train}})  / (y_{\text{gmm}}\& y_{\text{train}} \in T) \}
    \end{multline}
    $(X_{\text{key}}, Y_{\text{key}})$ will be used for both watermark insertion and verification. 
\end{enumerate}

This process can be reformulated in our unified framework as the DeepSigns's extraction function in the following form: 
   \begin{multline}
     \text{Ext}(M, K_{\text{ext}}) = \{\text{Ext}^i(M, K_{\text{ext}})\}_{i \in T} \\
     = \{\{f^l(x, w)\}_{x \in X_{\text{key}} \&  y_{\text{train}}=i}\}_{i \in T} 
   \end{multline}
where $\text{Ext}^i(M, K_{\text{ext}})$, with $i \in T$, is the set of activation maps of the input samples belonging to the trigger set $X_{\text{key}}$ with  $i$ as $y_{\text{gmm}}$ label. The extraction function of DeepSigns depends on the secret key $K_{\text{ext}}$ which corresponds to the trigger set. 

Using the trigger set, DeepSigns embeds in each of the secretly selected GMM classes a random binary sequence of $N$ bits. The watermark $b$ is thus defined as  $b = \{b_i\}_{i\in T}$, where   $b_i\in \{0, 1\}^{N}$ is inserted in the $i^{th}$ GMM class. 

This embedding process is similar to that of Uchida et al. scheme except that it uses $\{f^l(x,w)\}_{x \in X_{\text{key}}}$ instead of the vector of weights $\mathbf{w}$. Again, this insertion process can be reformulated with the help of a set of $s$ perceptrons as follows: Let us consider the $i^{th}$ secretly selected GMM class. We can define a single perceptron $\text{Proj}_{\theta_i}$ with Sigmoid as an activation function and $\theta_i$ as a set of parameters with as purpose the embedding of the watermark $b_i$.

 Each projection perceptron  $\text{Proj}_{\theta_i}$ can be seen as trained over the dataset $\{(A_j, b_{ij})\}_{j=1...N}$) where $A$ is a secret matrix $(K_{Proj}=A)$ of size $m\times N$, $A_j$ denotes the $j^{th}$ column of $A$ and $b_{ij}$ is the $j^{th}$ element of $b_i$. Note also that, with DeepSigns, all $\text{Proj}_{\theta_i}$ are trained over the same secret matrix $A$ but with different labels $b_i$. The distance used between $\text{Proj}_{\theta_i}(A)$ and $b_i$ is the cross entropy, considering a batch size equals to $N$. The parameters of $\text{Proj}_{\theta_i}$ are initialized with $\mu_l^i$ for $i\in T$ with the idea that the parameters which minimize the distance $d(\text{Proj}_{\theta_i}(A), b_i)$ stay close to the activation maps of the nonwatermarked model.

Based on the above statements, the watermarking regularization term of Deepsigns $E_{\text{wat}}$ can be reformulated for the first time such as:

\begin{equation}
\label{eq:ds_ewat}
E'_{wat} = \sum_{i\in T} (d(\text{Proj}_{\theta_i}(A), b_i) +  ||\theta_i - \underset{x \in X_{key} \& y_{train}=i}{\mathrm{mean}} (f^l(x, w^{wat})) ||)     
\end{equation}

To this term is added a second one, whose purpose is to prevent the parameters  $\{\theta_i\}_{i\in T}$ from converging to the means of the GMM classes that have not been selected for watermarking, i.e., $\{\mu_i^l\}_{i\not\in T}$. To do so, DeepSigns exploits a regularization term $E_{\text{sep}}$, see \eqref{eq:ds_esep}, to maximize their distance. 
\begin{equation}
\label{eq:ds_esep}
E_{\text{sep}} = \sum_{i\in T, j \not \in T} ||\theta_i - \mu_j^l ||    
\end{equation}

Therefore, the global DeepSigns loss function $E$ to embed the watermark and preserve the accuracy of the watermarked model is given by:
\begin{equation}
E = E_0(X_{\text{key}}, Y_{\text{key}}) + \lambda_1 E_{\text{wat}} - \lambda_2 E_{\text{sep}}
\end{equation}

\medskip
The watermark extraction process of DeepSigns stands on the two following steps: 
\begin{enumerate}
    \item Collect the activation maps corresponding to the selected trigger set $(X_{\text{key}}, Y_{\text{key}})$.
    \item Compute the statistical mean values of the activation maps of the selected trigger images. Take these mean values as an approximation of the parameters $\{\theta_i\}_{i\in T}$ where the watermarks $\{b_i\}_{i\in T}$ are assumed to be embedded. These mean values together with the owner’s private projection matrix $A$ are used to extract the watermarks $\{b^{ext}_i\}_{i\in T}$ following the procedure in \eqref{eq:extract_wat_DS}.
\end{enumerate}

\begin{equation} 
	b_i^{\text{ext}} = \text{HT}(\text{Proj}_{\underset{x \in X_{\text{key}} \& y_{\text{gmm}}=i}{\mathrm{mean}} (f^l(x, w^{\text{wat}}))}(A)) 
	\label{eq:extract_wat_DS}
\end{equation}

where HT is the Hard Thresholding operator that maps the values that are greater than 0.5 to 1 and the values smaller than 0.5 to 0. 

\paragraph*{Advantages and drawbacks of this approach :}
As defined, DeepSigns requires a trigger set, i.e., a secret subset of the training data, to embed (resp. extract) the watermark in (resp. from) the pdf of the activation maps associated with this trigger set. By doing so, it does not disturb too much the weights’ distribution without having to train a specific complex neural network such as RIGA. The security of DeepSigns is higher than \cite{uchida2017embedding, li2022encryption} and other static methods since it relies not only on the secret matrix $A$ but also on the trigger set ($X_{\text{key}}, Y_{\text{Key}}$) and on the secretly selected GMM classes to watermark. Another advantage of DeepSigns is that it is resistant to fine-tuning, pruning, and overwriting attacks but only for small insertion capacity. 

From \cite{rouhani2019deepsigns}, it is not clear if DeepSigns finetunes the target model only with the trigger set $(X_{\text{key}}, Y_{\text{key}})$ which represents $1\%$ of the training set. In case it is, the watermarked model may overfit the trigger set with as a consequence a loss of precision on the testing set. We experimentally noticed this accuracy drop. For a fair comparison with our proposal, we have assumed in the following experiments that to preserve the accuracy and to reach the performance of DeepSigns given in \cite{rouhani2019deepsigns}, the target model is fine-tuned over the entire training set and by changing the regularization terms depending on if the input data belongs to the trigger set or not. In other words, we trained the single perceptrons $\{\text{proj}_{\theta_i} \}_{i \in T}$ only on the activation maps of the trigger set data. And so, the embedding loss function can be written as: 
\begin{equation}
\label{eq:dp_new_loss}
E = E_0(X_{\text{Train}}, Y_{\text{Train}}) + \lambda_1 E_{\text{wat}} - \lambda_2 E_{\text{sep}}    
\end{equation}

In that way, we preserve well the precision of the watermarked model but at the price of a reduced insertion capacity. 
This can be explained by the fact that since the trigger set is a subset of the training set and represents only $1\%$ of the training set, it is preferable to classify the remaining $99\%$ of the data (which have a similar distribution to the trigger set data) than to minimize the distance between the trigger set activation maps and the parameters $\{\theta_i \}_{i \in T}$ in order to maintain the model's accuracy. To solve this problem, and as we will see in section \ref{sec:diction} devoted to our solution, we propose the use of a trigger set with a distribution different from the ones of the training and/or finetuning data. This solution allows the watermarked model both to insert a robust watermark and preserve its accuracy. 

The main weakness of DeepSigns stands in its insertion capacity. Increasing the length of the watermark will make it nonresistant to $FTA$ and $OWA$ attacks. To embed the watermark into the pdf distribution of DNN activation maps, DeepSigns takes advantage of the fact that DNN models possess nonconvex loss surfaces with many local minima that are likely to yield to very close accuracies \cite{choromanska2015loss}. Although using GMM clustering provides a reasonable approximation of the activation distribution obtained in hidden layer \cite{ioffe2015batch}, this one will necessarily be different for distinct local minima. If two data belong to the same class for a given local minimum, they are not necessarily in the same GMM class for other local minima. One can also question the interest in using clustering for large watermark capacities. Let us recall first that for DeepSigns the perceptrons or equivalently the projection functions $\{\text{Proj}_{\theta_i} \}_{i \in T}$ are trained to find the parameters $\{\theta_i \}_{i \in T}$ that minimize the distance $\{d(\text{Proj}_{\theta_i}(A), b_i)\}_{i \in T}$. Even though, $\{\theta_i \}_{i \in T}$ are initialized by the GMM means $\{\mu_l^i \}_{i \in T}$, one can see that there are no constraints in the loss $E_{\text{wat}}$ of DeepSigns to ensure $\{\theta_i \}_{i \in T}$ remain close or similar to $\{\mu_l^i \}_{i \in T}$. It is imposed that the watermarked means move away from the non-watermarked means $\{\mu_l^i \}_{i \not \in T}$. For large capacities, such a distance can be important. Therefore, there is not much interest in selecting the trigger set based on the GMM clustering since no constraints are applied on the parameters $\{\theta_i \}_{i \in T}$ and each local minima will provide a new clustering. Large distance between $\{\theta_i \}_{i \in T}$ and $\{\mu_l^i \}_{i \in T}$ may also cause a problem regarding the $1\%$ ratio between the trigger and training sets. A couple of finetuning epochs can be enough to bring closer $\{\theta_i \}_{i \in T}$ and $\{\mu_l^i \}_{i \in T}$ and erase the watermark. 

At least, another unclear point in \cite{rouhani2019deepsigns} is that Deepsigns activation maps $f_l(x, w)$ are said to be the outputs of activation functions like sigmoids or Relu, respectively defined as:
 \begin{align*}
     \sigma (x) & =  \frac{\mathrm{1} }{\mathrm{1} + e^{-x}} \\ 
     \text{Relu}(x) & = \max(0,x)
     \label{eq:relu}
 \end{align*}
Due to the fact, the parameters $\{\theta_i \}_{i \in T}$ that minimize the distance $\{d(\text{Proj}_{\theta_i}(A), b_i)\}_{i \in T}$ are vectors with positive or/and negative components (see above, $A$ is derived from a normal distribution $N(0,1)$ and $b$ is a random binary sequence) and that the outputs of the Sigmoid or Relu activation are always positive, there is a high risk the activation maps $\{\text{mean}_{x \in X_{\text{key}} \& y_{\text{train}}=i}(f_l(x, w))\}_{i \in T}$ do not converge to the parameters $\{\theta_i \}_{i \in T}$ (see $E_{\text{wat}}$, \eqref{eq:ds_ewat}). To avoid this problem and to retrieve the performance of DeepSigns given in \cite{rouhani2019deepsigns}, we decided to use the activation maps of the layer’s linear units, i.e., not the outputs of their activation function.

 \section{DICTION : a novel dynamic white box watermarking scheme for Deep neural networks}
 \label{sec:diction}

 \subsection{Architectural overview}
  DICTION, the solution we propose, is a novel dynamic scheme. It inserts a watermark into the activation maps of one or many layers concatenated of the target model. Its originality is twofold.  First of all, to ensure the fidelity requirement, i.e., not to disrupt the performance of the target model while embedding the watermark, it takes advantage of a trigger set issued from a distribution different from the training set.  Our idea is that if the trigger set is out of the distribution of the training data, e.g. data issued from a latent space defined as a Gaussian distribution as in the context of GAN generators \cite{goodfellow2014generative} and Variational AutoEncoder (VAE) decoders \cite{kingma2019introduction}, one will better preserve the pdf of the activation maps of the training samples, and consequently the accuracy of the target model, while increasing the robustness of the watermark.   

Second, to achieve a good trade-off between insertion capacity, and watermark robustness to PA, FTA, OA, and WKA, as well as to PIA, the projection function of DICTION is defined as a neural network that learns, using the trigger set, to map the activations of the target model $M^{\text{wat}}$ to a watermark $b$ and the activations of the non-watermarked model to a random watermark $b_r$. More clearly, such a projection function is comparable to the discriminator of a GAN model trained to distinguish, for samples in the latent space, the activation maps produced by a watermarked model from the ones of a non-watermarked model.

\begin{figure*}
     \centering
      \includegraphics[scale=.5]{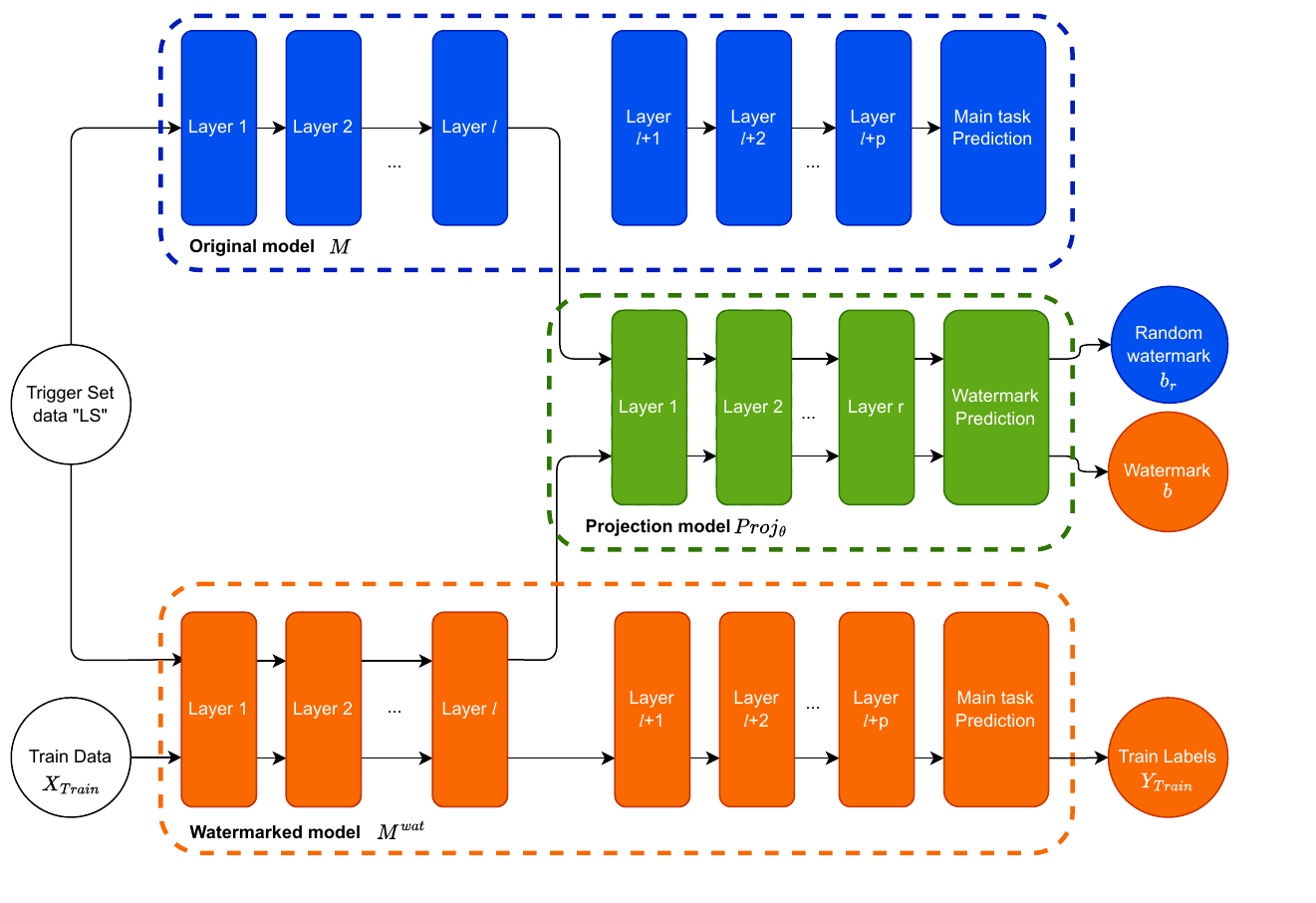}
 \caption{DICTION's workflow: The trigger set data "LS" is fed into both the original model $M$ and the watermarked model $M^{\text{wat}}$ to extract activation maps from the $l^{\text{th}}$ hidden layer. The projection function $\text{Proj}_{\theta}$, analogous to a discriminator, learns to map the activation maps of $M$ and $M^{\text{wat}}$ to a random binary string $b_r$ for the original model and to the watermark $b$ for the watermarked model. Concurrently, the watermarked model is trained on the training dataset $(X_{\text{Train}}, T_{\text{Train}})$ to preserve its performance on the primary task while embedding the watermark.}

     \label{fig:global_scheme}
 \end{figure*}

DICTION's workflow is illustrated in Fig. \ref{fig:global_scheme} and, as can be seen, in one training round, it considers the following steps to produce a watermarked model:

\begin{enumerate}

  \item We generate a trigger set image LS, of the same size as the training set images, from a latent space which follows a Gaussian distribution of mean ($\mu$) and standard deviation ($\sigma$)).  

  \item LS is then fed to the original or non-watermarked model $M$ and to the watermarked model under training $M^{\text{wat}}$ to compute, for a given layer $l$, the activation maps $f^l (\text{LS}, w^{\text{wat}})$ and $f^l (\text{LS}, \text{w})$, where: $w^{\text{wat}}$ are the weights of $M^{\text{wat}}$ and $w$ those of $M$, respectively. Notice that at the first round, $M^{wat}$ is initialized by the parameters of  $M$.

    \item $f^l (\text{LS}, \text{w})$ and $f^l (\text{LS}, w^{\text{wat}})$ are next used to train the projection model  $\text{Proj}_{\theta}^{\text{DNN}}$ assigning them the following labels:  the watermark $b$ for the activation maps of $M^{\text{wat}}$ and a random watermark $b_r$ for the ones of $M$.  This condition is implemented in our scheme to avoid that the training of $\text{Proj}_{\theta}^{\text{DNN}}$ only with $(f^l (\text{LS}, w^{\text{wat}}), b)$ lead to a trivial function $\text{Proj}_{\theta}^{\text{DNN}}$ that maps any of its input to $b$ (Integrity requirement - See section \ref{sec:intro}).

 \item To ensure the fidelity requirements, at each round, the target model $M^{\text{wat}}$ is trained also on the training set ($X_{\text{Train}}$, $Y_{\text{Train}}$). Regarding the original model, its parameters are frozen during all the watermarking session.  

\end{enumerate}
These steps are repeated until the BER between the extracted watermark $b^{\text{ext}}$ from $M^{\text{wat}}$ and the embedded watermark $b$ is equal to $0$ while maintaining a good accuracy of the target model $M^{\text{wat}}$. 

To sum up, our embedding process is similar to the training of a GAN, where the projection model plays the role of the discriminator that learns to classify the activation maps of the original model $M$ as $b_r$ and the ones of the target model  $M^{\text{wat}}$ as $b$. The first $l$ layers of the target model $M^{\text{wat}}$ play the role of the generator and they are trained to generate appropriate activation maps for the discriminator as well as for the main task prediction (i.e. classification). The original model $M$ is not trained during the embedding process and its activation maps are fed to the projection model to prevent it from becoming a trivial function. Note that, the watermarked model $M^{\text{wat}}$ and the projection model $\text{Proj}_{\theta}^{\text{DNN}}$ are trained simultaneously for fast convergence.

\subsection{Formal definition of DICTION }
Let us formulate our scheme as we did in Section \ref{subsec: whiteWat} for the state-of-the-art white box schemes, through its watermarking and embedding regularization terms: $E_{wat}$ and $E$, respectively. Let us consider a trigger set defined as latent space with a normal distribution $N(\mu, \sigma)$, of mean $(\mu)$ and of standard deviation $(\sigma)$. From this standpoint, the feature extraction function for a given target model $M$ is such as: 
\begin{equation}
 \text{Ext}(M, K_{\text{ext}}) = f^l(\text{LS}, w)*Z    
\end{equation}
where:  $K_{\text{ext}} = \{l, Z, \text{LS}\}$ is the secret extraction key composed with $l$ the index of the layer to be watermarked and $Z$ a permutation matrix that is used to secretly select a subset of features from $f^l(\text{LS}, w)$ and secretly order them.  $\text{LS}$ is a sample from the latent space (i.e. an image the pixel values of which follow a normal distribution); 

As stated previously and as shown in Fig. \ref{fig:global_scheme}, our projection function $\text{Proj}_{\theta}^{\text{DNN}}$ is a neural network. It takes as input the extracted features $\text{Ext}(M^{\text{wat}}, K_{\text{ext}})$ to output the watermark $b$. The sizes of its input and output are thus equal to the size of extracted features, $|\text{Ext}(M,  K_{\text{ext}})|$, and to the watermark size $|b|$, respectively. To embed $b$ in the secretly selected layer $l$ of the target model, the watermarking regularization term of our scheme is defined as: 
\begin{multline}
\label{eq:wat_diction}
   E_{\text{wat}} = d(\text{Proj}_{\theta}^{\text{DNN}}(\text{Ext}(M^{\text{wat}},  K_{\text{ext}})), b)  + \\ d(\text{Proj}_{\theta}^{\text{DNN}}(\text{Ext}(M, K_{\text{ext}})), b_r)  
\end{multline}
where: $M^{\text{wat}}$ is the watermark version of $M$; $b_r$ is a random watermark; $d$ is a watermark distance measure.  $b_r$ is used to ensure that $\text{Proj}_{\theta}^{\text{DNN}}$ projects the desired watermark $b$ only for $M^{\text{wat}}$. Note that $d$ depends on the type of the watermark. For instance, in the case $b$ is a binary sequence, one can use cross-entropy whereas in the case it is an image, the pixel mean squared error is more relevant. Our global embedding loss is given as :

\begin{equation}
\label{eq:embed_diction}
E = E_0(X_{\text{Train}}, Y_{\text{Train}}) + \lambda E_{\text{wat}}
\end{equation}
where: $\lambda$ is here to adjust the tradeoff between the original target model loss term $E_0$ and the watermarking regularization term $E_{\text{wat}}$;  $(X_{\text{Train}}, Y_{\text{Train}})$ is the training set. As it can be seen, $E_{\text{wat}}$ does not depend on  $(X_{\text{Train}}, Y_{\text{Train}})$ but on the trigger set only. More clearly, the parameters of $M^{wat}$ are updated accordingly to the following regularization term from $E$: 

\begin{multline}
\label{eq:diction_emb_w}
     E_{\mathbf{w}^{\text{wat}}} = E_0(X_{\text{Train}}, Y_{\text{Train}}) + \\ \lambda d(\text{Proj}_{\theta}(\text{Ext}(M^{\text{wat}}, K_{\text{ext}})), b)
\end{multline}

By doing so, the parameters $\mathbf{w}_{\text{wat}}$ of $M^{\text{wat}}$ are updated such that they keep the accuracy of the original target model and minimize the distance between the projections of the activation maps and the watermark $b$. On their side, the parameters of $\text{Proj}_{\theta}^{DNN}$  are updated based on the following regularization terms in $E$: 
\begin{multline}
\label{eq:diction_emb_teta}
 E_{\theta} = \lambda( d(\text{Proj}_{\theta}^{\text{DNN}}(\text{Ext}(M^{\text{wat}}, K_{\text{ext}})), b) \\  + d(\text{Proj}_{\theta}(\text{Ext}(M, K_{\text{ext}})), b_r))
\end{multline}

 As previously mentioned, the projection model $\text{Proj}_{\theta}^{\text{DNN}}$ serves as a discriminator that associates the projection of the trigger set activation maps of $M^{\text{wat}}$ to $b$ and those of $M$ to $b_r$. One can notice from \eqref{eq:diction_emb_w} and \eqref{eq:diction_emb_teta} that the term $d(\text{Proj}_{\theta}(\text{Ext}(M^{\text{wat}}, K_{\text{ext}})), b)$ is shared in the update of $M^{\text{wat}}$ and $\text{Proj}_{\theta}^{DNN}$. This enables the optimal compromise between the parameters $w^{\text{wat}}$ and $\theta$ so that the projection of the activation maps is close to $b$, with no impact on the accuracy, and is unique for $M^{\text{wat}}$. In practice, the parameters of $M^{\text{wat}}$ and $\text{Proj}_{\theta}^{\text{DNN}}$ are trained alongside the embedding process to enable fast convergence.

 The watermark detection process works as follows: Let us consider a suspicious watermarked model $M^*$. Based on the features extraction key $K_{\text{ext}} = \{ l, Z,  \text{LS}\}$, the projection key $K_{\text{proj}} = \text{Proj}_{\theta}^{DNN}$ and the latent space $N(\mu, \sigma)$ , the watermark extraction is such as :
 \begin{equation}
     b^{\text{ext}} =  \text{HT}(\underset{\text{LS} \in N(\mu, \sigma) }{\mathrm{mean}} \text{proj}_{\theta}^{\text{DNN}}(\text{Ext}(M^{\text{wat}},  K_{\text{ext}}))
 \label{eq:extract_wat_diction}
 \end{equation}
 where HT is the Hard Thresholding operator at $0.5$.  In contrast to a fixed subset of the training set as used in DeepSigns (see \eqref{eq:extract_wat_DS}), in DICTION extraction, you can sample an endless number of latent space images.  By doing so, the watermarked model and projection function are extremely resistant to attacks such as fine-tuning, pruning, and overwriting.  
 
 This is because the watermarked model and projection function overfit the latent space rather than a few images. In addition, it decreases the storage complexity of the detection as one just has to record the latent space parameters (i.e. its mean and standard deviation) rather than storing a trigger constituted of images, the number and size of which might be quite big in some applications.  Notice also that the use of a latent space as a trigger set allows for preserving the distribution of weights and activation maps of the watermarked model (see Fig. \ref{fig:detection_attack}). Moreover, it avoids the use of another neural network like $F_{det}$ in RIGA scheme which needs to find a good hyperparameterization (e.g., clamping weights -see Section \ref{sec:related_work}) and increasing thus the computational complexity of the watermarking algorithm.

\section{Experimental Results}
\label{sec:experiments}

We evaluated DICTION in terms of its: impact on model performance (Fidelity); robustness against three watermark removal techniques (overwriting, fine-tuning, and weight pruning); and, resilience to watermark detection attacks (Security). For all attacks, we considered the most hard threat model, assuming that the attacker has access to, or knowledge of, the training data and the watermarked layer.

\subsection{Datasets and Models}
The following experiments have been conducted on two well-known public benchmark image datasets in the context of image classification: CIFAR-10 \cite{krizhevsky2014cifar} and MNIST \cite{lecun1998gradient}. In addition, we have utilized four distinct neural network architectures as target models: three have been experimented with DeepSigns \cite{rouhani2019deepsigns} - MLP, CNN and ResNet-18, and one with RIGA \cite{wang2019riga} - LeNet. Table \ref{tab:model_bench} summarizes their topologies depending on the target dataset, the number of epochs, and the corresponding achieved accuracies. The implementations of DeepSigns \cite{rouhani2019deepsigns}, ResEncrypt \cite{li2022encryption}, and DICTION, along with their configurations, are publicly available at \url{https://github.com/Bellafqira/DICTION}.

\begin{table}
\caption{Benchmark neural networks: architectures, number of epochs, and classification baseline accuracies on two public datasets (CIFAR10 \cite{krizhevsky2014cifar} and MNIST \cite{lecun1998gradient}). In the models, `64C3(1)` denotes a convolutional layer with 64 output channels and $3 \times 3$ filters applied with a stride of 1; `MP2(1)` indicates a max-pooling layer over $2 \times 2$ regions with a stride of 1; `512FC` refers to a fully connected layer with 512 output neurons; `BN` stands for batch normalization. Activation functions ReLU and Sigmoid are used across all benchmarks.}

\centering
\begin{tabular}{|c|c|c|c|c|p{3cm}|}
\hline
\textbf{Benchmark ID} & \textbf{Dataset} & \textbf{Baseline Acc} & \textbf{Number of Epochs} & \textbf{DL Model Type} & \textbf{DL Model Architecture} \\
\hline 
1 & MNIST & 98.06\% & 50 & MLP & 784-512FC-512FC-10FC \\
\hline
2 & CIFAR10 & 87.17\% & 100 & CNN & 3*32*32-32C3(1)-32C3(1)-MP2(1)-64C3(1)-64C3(1)-MP2(1)-512FC-10FC \\
\hline
3 & CIFAR10 & 91.7\% & 200 & ResNet18 & Please refer to \cite{he2016deep} \\
\hline
4 & MNIST & 99.16\% & 50 & LeNet & 1*28*28-24C3(1)-BN-24C3(1)-BN-128FC-64FC-10FC \\
\hline
\end{tabular}
\label{tab:model_bench}
\end{table}

\subsection{Watermarking Settings}

In all our experiments, a watermark $b$ of 256 bits is embedded in the second-to-last layer of the target model, following the approach used in DeepSigns. This watermark could, for example, be the hash of the model owner's identifier generated using the SHA-256 hash function. We use the bit error rate (BER) to measure the discrepancy between the original and extracted watermarks. The BER is defined as:

\begin{equation}
    \text{BER}(b^{\text{ext}}, b) = \frac{1}{n} \sum_{i=1}^n I(b^{\text{ext}}_i \neq b_i)
\end{equation}
where $b^{\text{ext}}$ is the extracted watermark, $I$ is the indicator function which returns 1 if the condition is true and 0 otherwise, and $n$ is the size of the watermark ($n=256$). A BER close to 0 indicates that the two watermarks are identical, while a BER close to 0.5 means that the watermarks are not correlated.

We used a simple 2-layer fully connected neural network as the architecture for the DICTION projection function $\text{Proj}_{\theta}$ in our experiments. The normal distribution $N(\mu,\sigma)=N(0,1)$ serves as a latent space to embed the watermark. and the selection matrix $Z$ as the identity. We use Adam optimizer with the learning rate $10^{-3}$, batch size $100$, and weight decay $10^{-4}$ to train the DICTION projection function. We set $\lambda = 1$ in  \eqref{eq:embed_diction}. We used 10 epochs for each benchmark to embed the watermark.

Regarding the implementation of DeepSigns \cite{rouhani2019deepsigns}, we used also the second-to-last layer for embedding, as in its original version. $\lambda_1$ and $\lambda_2$ have been set to $0.01$ to obtain a BER equal to zero.

The number of watermarked classes $s$ equals $2$ and $N$ to $128$ (i.e., a watermark of 256 bits), and the watermark projection matrix $A$ is generated based on the standard normal distribution $N(0,1)$.  The trigger set corresponds to $1\%$ of the training data. 

For ResEncrypt, we used 2-layer fully connected neural networks as architecture for Mapping Net. We recall that the role of the Mapping Net is to expand the size of the selected weights to the watermark. In our implementation, we used an expansion factor equal to 2 for LeNet and 1 for the other benchmarks. 

\subsection {Fidelity} 
By definition, the accuracy of the watermarked model shall not be degraded compared to that of the target model. Table \ref{tab:model_bench} summarizes the original models' accuracy and Table \ref{tab:model_embed} the accuracy of watermarked models after the embedding of a watermark of 256 bits. As can be seen, the accuracies of the watermarked models are very close to those of the non-watermarked models for DeepSigns, ResEncrypt, and DICTION. This means that they simultaneously optimize the accuracy of the underlying model, as well as minimize the watermark loss term ($E_{\text{wat}}$) as discussed in Section \ref{sec:diction}. %In some cases, we even observe a slight accuracy improvement of DICTION compared to the baseline. Such an improvement is mainly due to the fact that the additive loss functions ($E_{\text{wat}}$) act as a form of a regularizer during the training phase of the original model. Regularization, in turn, helps the model avoid over-fitting by inducing a small amount of noise in the model.

\begin{table*}
    \caption{BER and accuracy of DeepSigns, ResEncrypt, and DICTION for the four benchmarks after embedding a watermark of 256 bits}
    \centering
    \begin{tabular}{|c|c|c|c|c|c|c|c|}
    \hline
   \multirow{2}{*}{ \textbf{Benchmark ID} }     & \multirow{2}{*}{\textbf{Baseline Acc} } &  \multicolumn{2}{c|}{\textbf{DeepSigns} }  & \multicolumn{2}{c|}{\textbf{DICTION}} & \multicolumn{2}{c|}{\textbf{ResEncrypt}} \\
\cline{3-8}
        & & BER & Model accuracy & BER & Model accuracy  & BER & Model accuracy\\

    \hline
                1 & 98.06\%  & 0 & 97.73\% & 0 & 98.06\% & 0 & 97.57\%\\
    \hline
                2 & 87.17\% & 0 & 85.18\% & 0 & 85.65\% & 0 & 84.98\% \\
    \hline
                3 & 91.7\% & 0 & 91.36\% & 0 & 91.58\% & 0 & 90.84\% \\
    \hline
                4 & 99.16\% & 0 & 98.75\% & 0 & 99.1\% & 0 & 98.68\%\\
    \hline
    \end{tabular}

    \label{tab:model_embed}
\end{table*}

\subsection {Robustness} 
We evaluate the robustness of our scheme against three contemporary removal attacks as discussed in Section \ref{sec:intro}. 

\begin{figure*}
     \centering
     \begin{subfigure}[b]{\textwidth}
         \centering
         \includegraphics[scale=0.25]{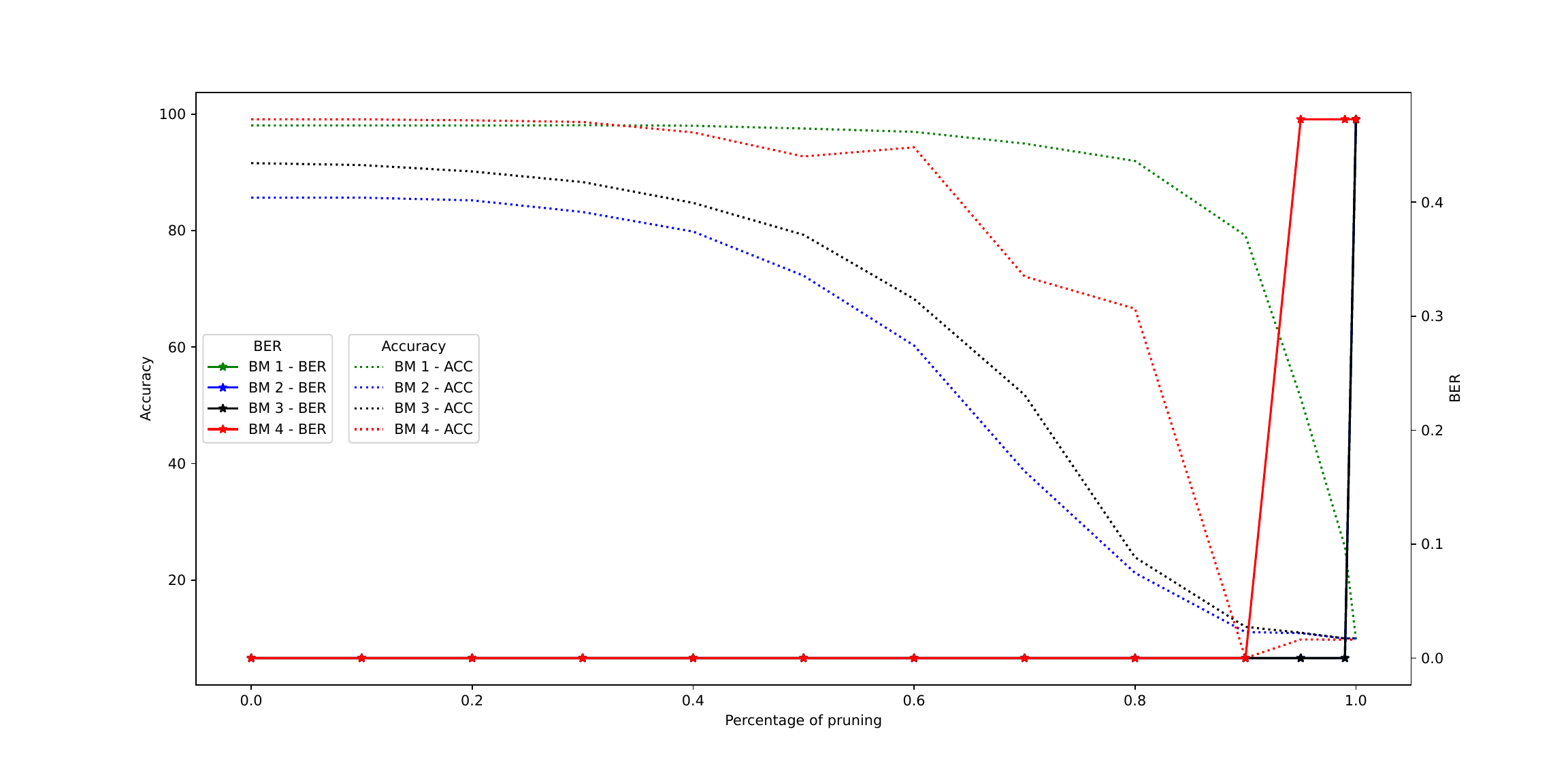}
         \caption{ DICTION  }
         \label{fig:prun_diction}
     \end{subfigure}
     \hfill
     \begin{subfigure}[b]{\textwidth}
         \centering
         \includegraphics[scale=0.25]{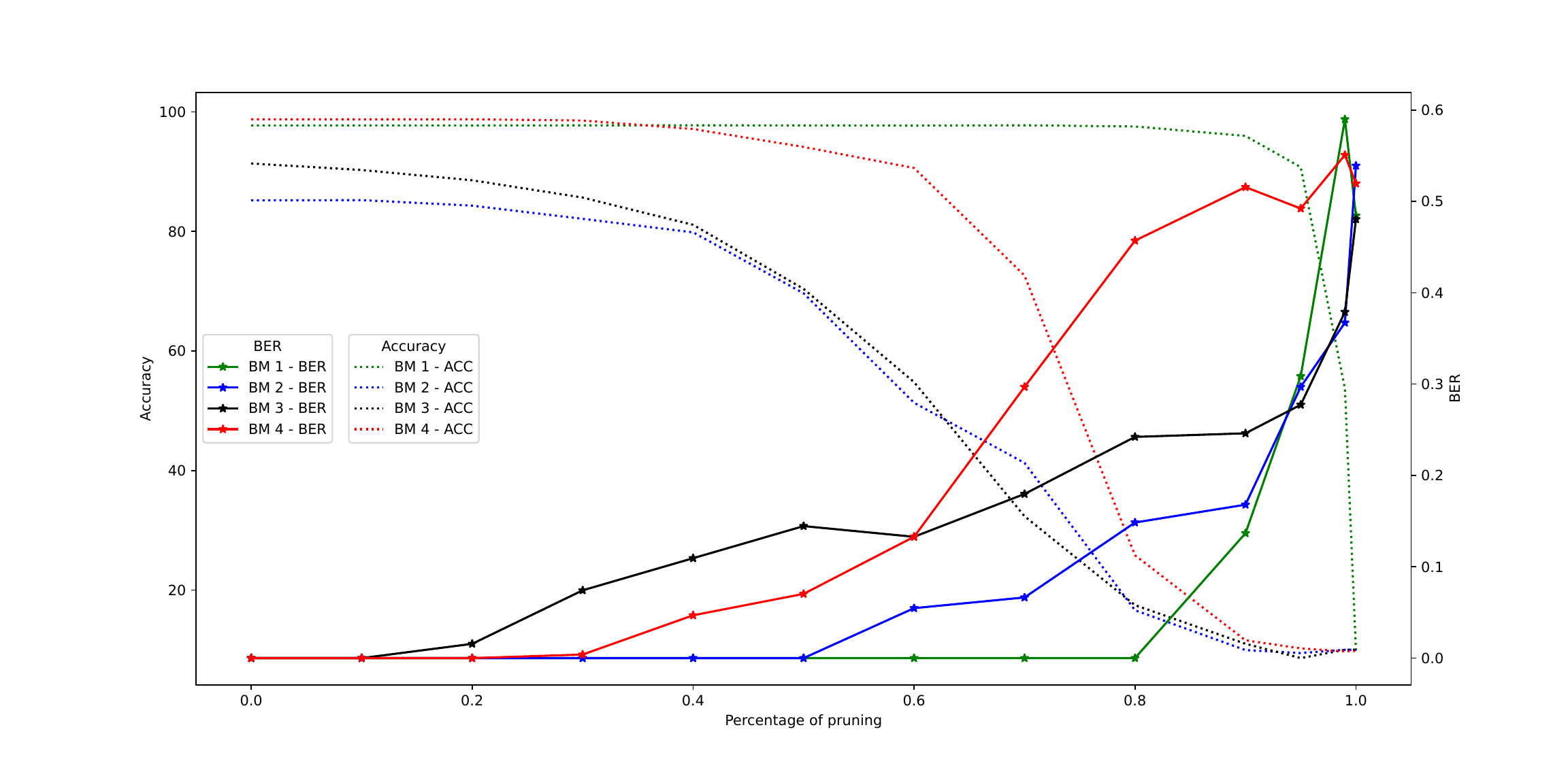}
         \caption{DeepSigns}
         \label{fig:prun_deepsigns}
     \end{subfigure}
    \hfill
    \begin{subfigure}[b]{\textwidth}
        \centering
         \includegraphics[scale=0.25]{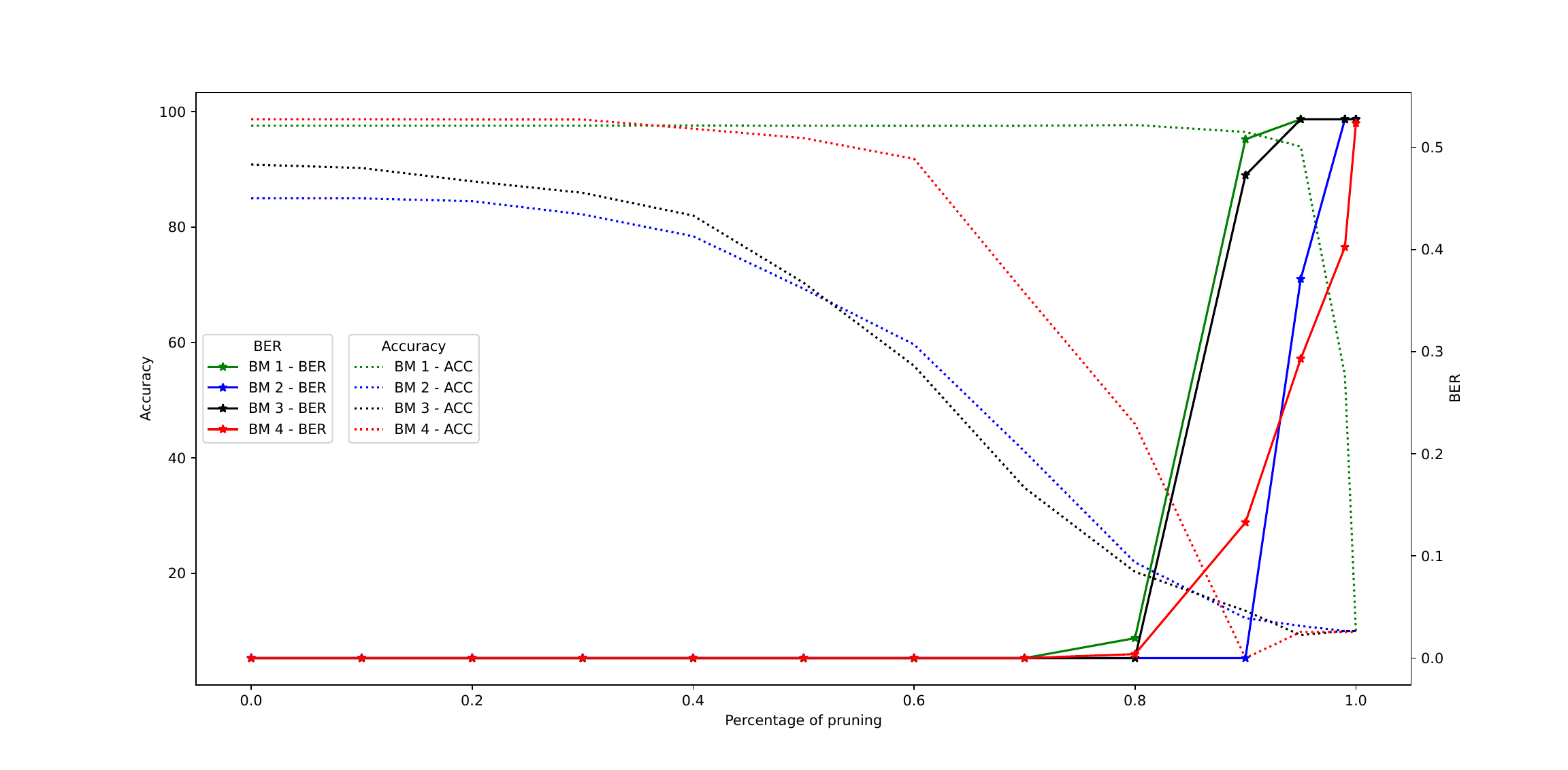}
         \caption{ResEncrypt}
         \label{fig:prun_resEncrypt}
    \end{subfigure}

        \caption{Robustness of DICTION (a), DeepSigns (b)  and ResEncrypt (C) against the pruning attack for the four benchmark models (BM 1 to 4 as given in Table \ref{tab:model_bench}). Solid lines correspond to BER and dashed lines to the accuracy of the attacked watermarked models depending on the pruning percentage "$\alpha$".
        }

        \label{fig:pruning} 
\end{figure*}

\vspace{0.7em}
\noindent \textbf{Pruning attack.} We use the pruning approach proposed in~\cite{han2015learning} to compress the watermarked models. For each layer of a watermarked model, this one first sets $\alpha\%$ of the weights of the lowest valuer to zero. Fig. \ref{fig:pruning}(a) and (b) illustrate the impact of pruning on watermark extraction/detection as well as on model accuracy depending on different values of $\alpha$. As it can be seen, DICTION  can tolerate a pruning up to $90\%$ and $95\%$ and up to $80\% $ for ResEncrypt whatever the benchmark networks. One can also note that DeepSigns is more vulnerable when a 256-bit watermark is inserted. Moreover, when pruning yields a substantial bit error rate (BER) value, the attacked watermarked model suffers from a large accuracy loss compared to the baseline. As such, one cannot remove with such an attack the embedded watermark without loss of accuracy compared to the baseline. Looking at these two dimensions, DICTION offers better performance. 

\begin{table}
    \caption{Robustness of DICTION ResEncrypt and DeepSigns against the model fine-tuning attack.}
    \centering
    \resizebox{\textwidth}{!}{%
    \begin{tabular}{|c|cc|cc|cc|cc|}
    \hline
        \textbf{Benchmarks} & \multicolumn{2}{c|}{\textbf{BM 1}} & \multicolumn{2}{c|}{\textbf{BM 2}} & \multicolumn{2}{c|}{\textbf{BM 3}} & 
        \multicolumn{2}{c|}{\textbf{BM 4}} \\
        \hline
        \textbf{Epochs} & 50 & 100 & 50 & 100 & 50 & 100 & 50 & 100 \\
        \hline
        \textbf{ACC DICTION} & 98.27\% & 98.3\% & 86.81\% & 86.69\% & 91.59\% & 91.58\% & 99.17\% & 99.16\% \\
        \hline
        \textbf{BER DICTION} & 0 & 0 & 0 & 0 & 0 & 0 & 0 & 0 \\
        \hline
        \textbf{ACC ResEncrypt} & 98.54\% & 98.64\% & 87.37\% & 87.3\% & 91.31\% & 90.98\% & 99.17\% & 99.2\% \\        
        \hline
        \textbf{BER ResEncrypt} & 0 & 0 & 0 & 0 & 0 & 0 & 0.5234 & 0.5273 \\
        \hline
        \textbf{ACC DeepSigns} & 98.53\% & 97.83\% & 86.9\% & 87.49\% & 90.89\% & 91.48\% & 99.14\% & 99.14\% \\
        \hline
        \textbf{BER DeepSigns} & 0.0977 & 0.1445 & 0 & 0.0234 & 0.1406 & 0.1641 & 0.3398 & 0.3320 \\
    \hline
    \end{tabular}%
    }
    \label{tab:fine_tuning}
\end{table}

\noindent \textbf{Model fine-tuning.} Fine-tuning is another form of transformation attack that a third-party might use to remove the watermark. In its more efficient implementation, it is assumed that the watermarked model is fine-tuned using the original training data and the loss of the target model, i.e., the conventional cross-entropy loss function, in our case. More clearly, the attacker does not exploit the watermarking loss functions $E_{\text{wat}}$. Note that fine-tuning deep learning models makes the underlying neural network converge to another local minimum that is not necessarily equivalent to the original one in terms of prediction accuracy. In our experiments, the fine-tuning attack is performed by multiplying the last learning rate used at the training stage by a factor of 10 and then we divide it by 10 after 20 epochs. This choice allows the model to look for a new local minima. Table~\ref{tab:fine_tuning}, summarizes the impact of fine-tuning on the watermark detection rate expressed in BER for all benchmark models. As shown, DICTION can successfully detect the watermark even after fine-tuning the deep neural network for many epochs. That is not the case with DeepSigns or ResEncrypt.  Resencrypt's results can be explained by the fact that in BM 4, the number of parameters in the watermarked  layer is small, unlike in benchmarks 1, 2 and 3. DeepSigns, on the other hand, has a limited capacity, so inserting a 256-bit watermark weakens its robustness. 

\begin{table*}
  \caption{BER of DICTION, DeepSigns and ResEncrypt for the four benchmarks after overwriting the watermarked model with a watermark of 256 bits or 512 bits using different keys}
    \centering
    \begin{tabular}{|c|c|c|c|c|c|c|}
    \hline
  \multirow{2}{*}{\textbf{Benchmark ID} }      &  \multicolumn{2}{c|}{\textbf{DICTION}}  & \multicolumn{2}{c|}{\textbf{DeepSigns}}  & \multicolumn{2}{c|}{\textbf{ResEncrypt}} \\
\cline{2-7}
         & 256 & 512 & 256 & 512 & 256 & 512 \\

    \hline
                1 & 0 & 0 & 0.19921875 & 0.2773 & 0.52734375 & 0.52734375 \\
    \hline
                2 & 0 & 0 & 0.3671875 & 0.41796875  & 0 & 0 \\
    \hline
                3 & 0 & 0 & 0.390625 &  0.4375\% & 0 &  0\\
    \hline
                    4 & 0 & 0 &  0.484375  &0.484375  &  0.40625 & 0.41796875 \\
    \hline
    \end{tabular}
  
    \label{tab:overweiting}
\end{table*}
\vspace{0.7em}
\noindent \textbf{Watermark overwriting.} Assuming the attacker is aware of the watermarking technique, he may attempt to damage the original watermark by embedding a new watermark in the watermarked model. In practice, the attacker does not have any knowledge about the location of the watermarked layers. However, in these experiments, considering the worst-case scenario, we assumed that the attacker knows where the watermark is embedded but does not know the original watermark nor the projection function. To perform the overwriting attack, the attacker follows the protocol discussed in Section \ref{sec:diction} to embed a new watermark of the same size as the original watermark (i.e., a 256-bit message) or of a larger size (i.e. a 512-bit message), using different or same latent space. Table \ref{tab:overweiting} summarizes the results we obtained against this attack for all four benchmarks. As shown, DICTION is robust against the overwriting attack and can successfully detect the original embedded watermark in the overwritten model while DeepSigns and  ResEncrypt, the  watermarks are perturbed and sometimes are  completely erased (achieving a BER close to 50\% in most cases) especially with shallower models such as BM1 and BM4. 
% I need to explain what an efficient fine tuning attack : 
%First doesnt require the same amount of data of the training set and epochs as for training a model from scratch
%
\subsection {Watermark Detection Attack (Security requirement).} 

\begin{figure*}
     \centering
     \begin{subfigure}[b]{0.4\textwidth}
         \centering
         \includegraphics[scale=0.18]{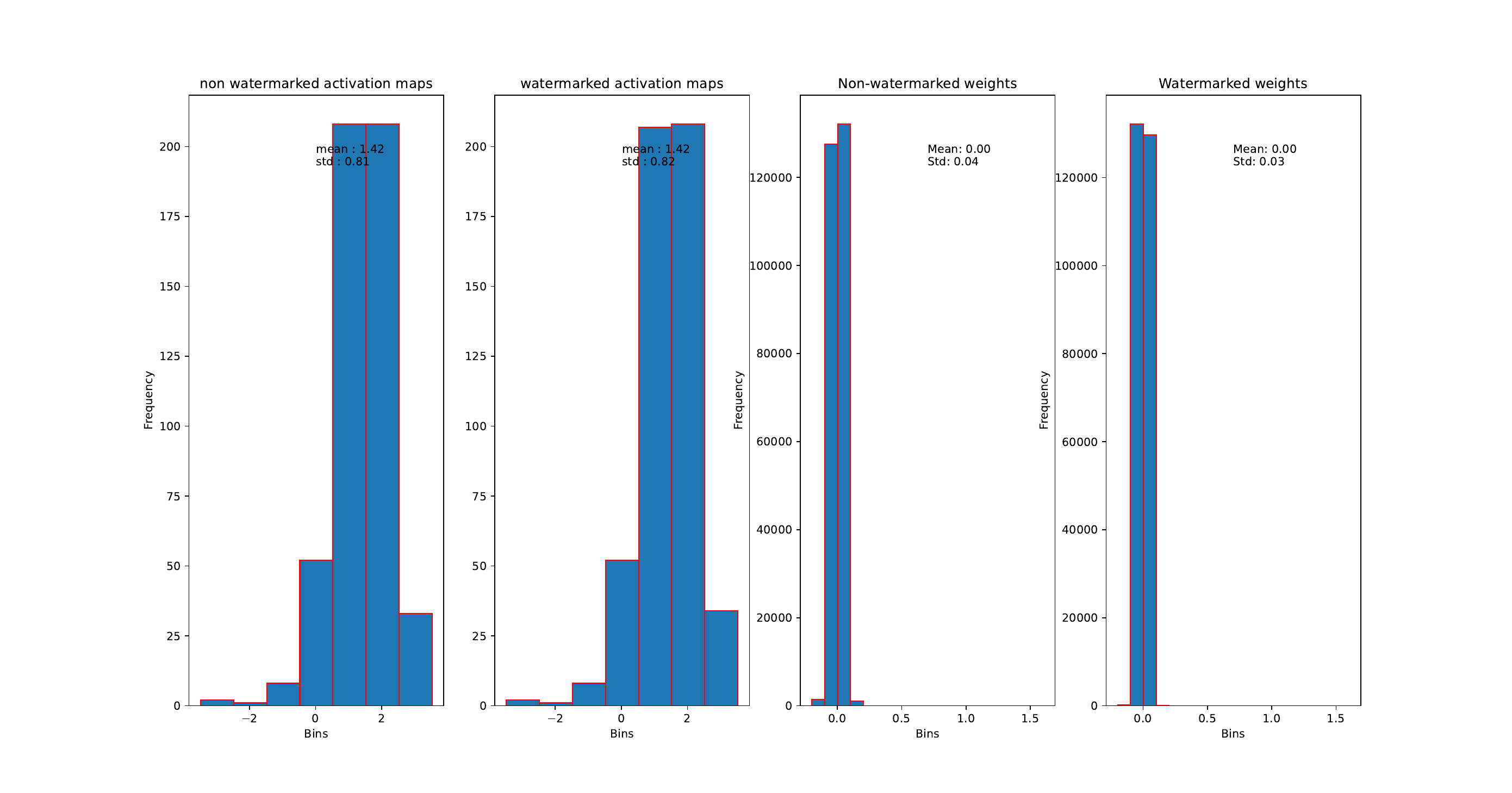}
         \caption{MLP}
         \label{fig:dists_mlp}
     \end{subfigure}
     \hfill 
     \begin{subfigure}[b]{0.5\textwidth}
         \centering
         \includegraphics[scale=0.18]{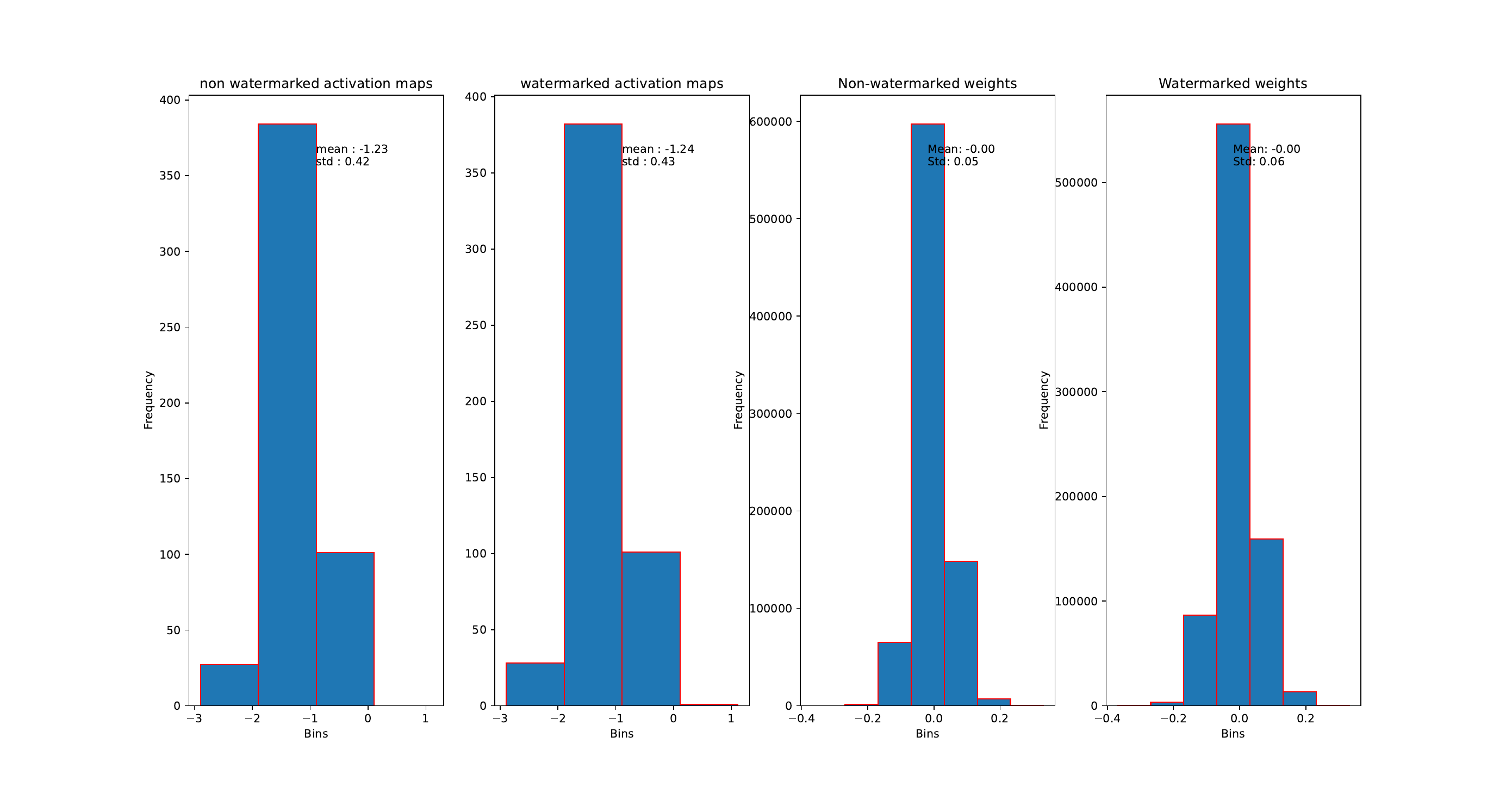}
         \caption{CNN}
         \label{fig:dists_cnn}
     \end{subfigure}
     \newline
     \begin{subfigure}[b]{0.4\textwidth}
         \centering
         \includegraphics[scale=0.18]{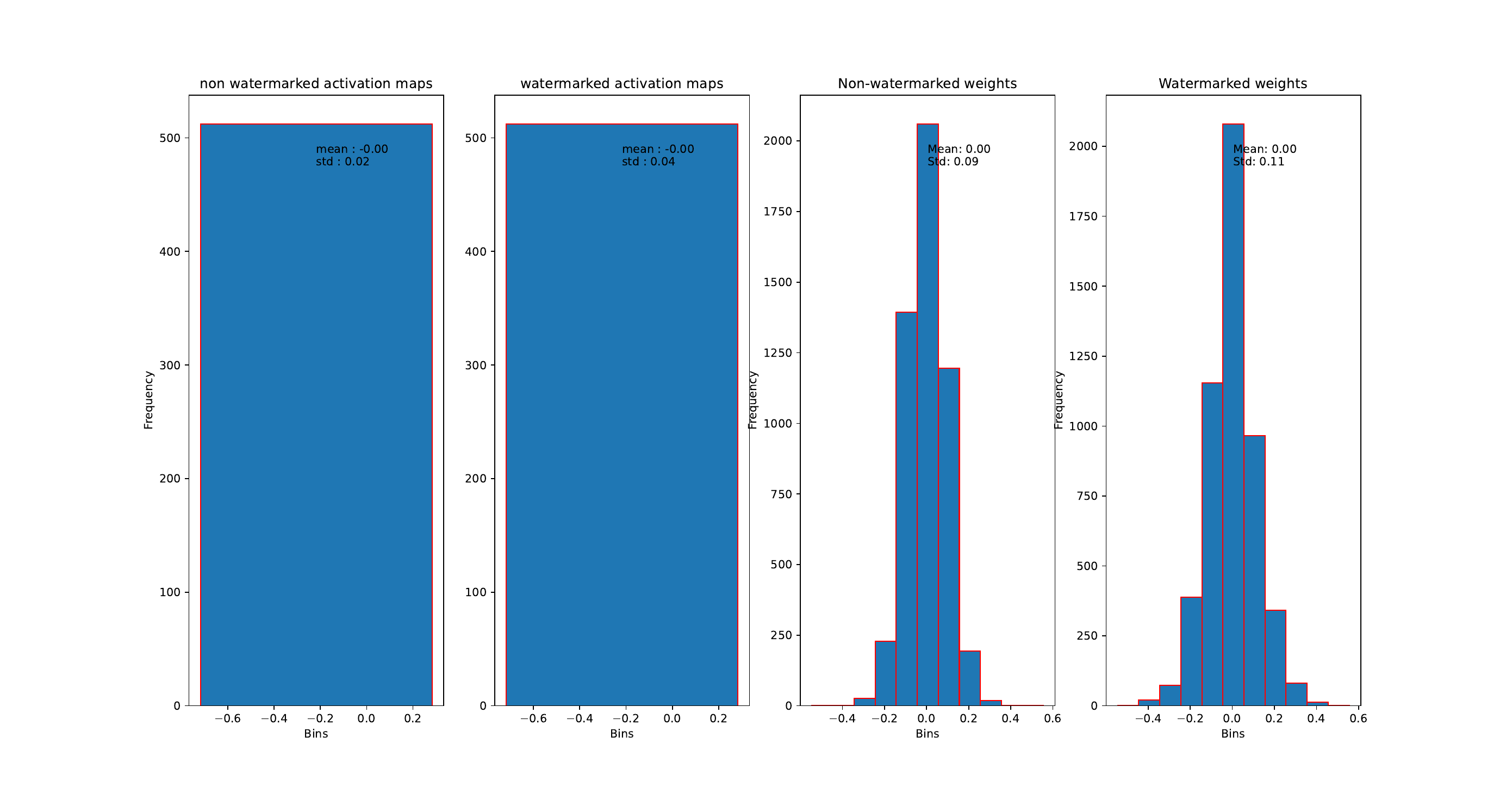}
         \caption{Resnet18}
         \label{fig:dists_resnet18}
     \end{subfigure}
         \hfill
    \begin{subfigure}[b]{0.5\textwidth}
         \centering
         \includegraphics[scale=0.18]{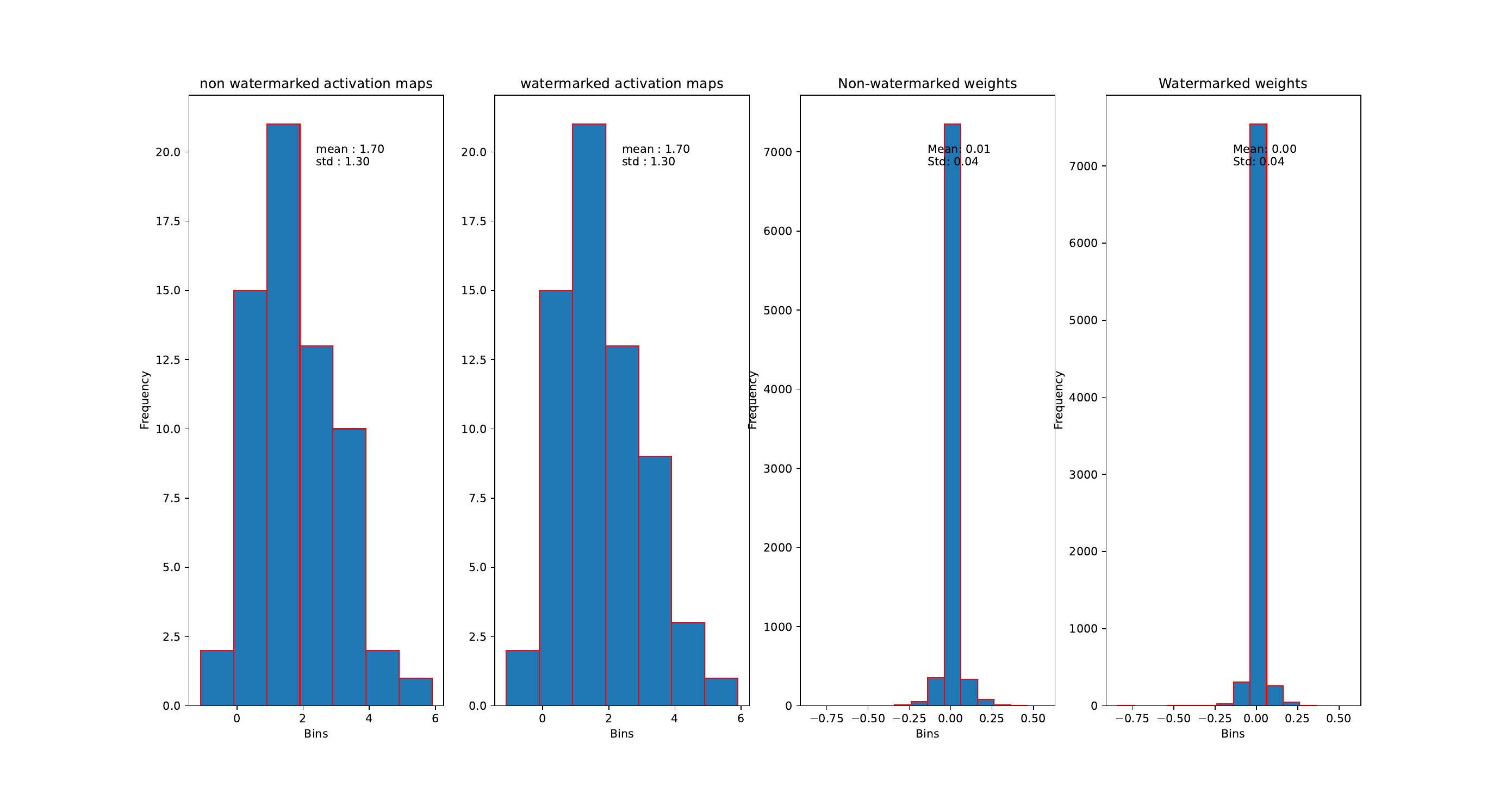}
         \caption{LeNet}
         \label{fig:dists_mlp_riga}
     \end{subfigure}
         \caption{DICTION: Distribution of the activation maps and the weights of the watermarked and non watermarked models considering the four benchmarks. }
        \label{fig:detection_attack} 
\end{figure*}

We tested all schemes against the attack proposed by Wang et al. \cite{TWang} and PIA. The embedding of the watermark should not leave noticeable changes in the probability distribution spanned of the watermarked model. Fig. \ref{fig:detection_attack} illustrates the histogram of the activation maps of the layer of the watermarked model and of the same layer from the non-watermarked one. DICTION preserves the distribution spanned by the model while robustly embedding the watermark. Note that the range of activation maps is not deterministic in the different models and cannot be used by malicious users to detect the existence of a watermark.  Table \ref{tab:mean_std_dist} gives the mean and standard deviation of the parameter distribution and activation maps of the watermarked layer for the fifth watermarking schemes. We can see that Diction least disturbs the distribution of the  model parameters for all four benchmarks. 

%\begin{figure*}
%     \centering
%     \begin{subfigure}[b]{0.4\textwidth}
%         \centering
%         \includegraphics[scale=0.2]{images/dists/dist_resencrypt_mlp.pdf}
%         \caption{MLP}
%         \label{fig:dists_mlp}
%     \end{subfigure}
%     \hfill 
%     \begin{subfigure}[b]{0.5\textwidth}
%         \centering
%         \includegraphics[scale=0.2]{images/dists/dist_resencrypt_cnn.pdf}
%         \caption{CNN}
%         \label{fig:dists_cnn}
%     \end{subfigure}
%     \newline
%     \begin{subfigure}[b]{0.4\textwidth}
%         \centering
%         \includegraphics[scale=0.2]{images/dists/dist_resencrypt_resnet18.pdf}
%         \caption{Resnet18}
%         \label{fig:dists_resnet18}
%     \end{subfigure}
%         \hfill
%    \begin{subfigure}[b]{0.5\textwidth}
%         \centering
%         \includegraphics[scale=0.2]{images/dists/dist_resencrypt_mlp_riga.pdf}
%%         \caption{LeNet}
 %        \label{fig:dists_mlp_riga}
 %    \end{subfigure}
 %        \caption{ResEncrypt: Distribution of the activation maps and the weights of the watermarked and non watermarked models considering the four %benchmarks. }
 %       \label{fig:detection_attack_ResEncrypt} 
%\end{figure*}

\begin{table}
    \centering
    \caption{Mean and std of the weights and activation maps distribution of the baseline model and the watermarked models}
    \resizebox{\textwidth}{!}{%
    \begin{tabular}{|c|c|c|c|c|c|c|c|c|}
    \hline
          & \multicolumn{2}{c|}{BM1} & \multicolumn{2}{c|}{BM2} & \multicolumn{2}{c|}{BM3} & \multicolumn{2}{c|}{BM4} \\
    \cline{2-9}
          & Act\_maps & Weights & Act\_maps & Weights & Act\_maps & Weights & Act\_maps & Weights \\
    \hline
    Baseline    &  \textbf{1.42 $\pm$ 0.81}& \textbf{0 $\pm$ 0.04}  &   \textbf{-1.23 $\pm$ 0.42}  & \textbf{0 $\pm$ 0.05}  &  \textbf{0 $\pm$ 0.02}  & \textbf{0 $\pm$ 0.09}  & \textbf{1.70 $\pm$ 1.30}  & \textbf{0.01 $\pm$ 0.04}  \\
    \hline
    ResEncrypt  & 0 $\pm$ 0.53 & -0.01 $\pm$ 0.02 & -1.64 $\pm$ 0.45 & \textbf{0}  $\pm$ 0.06 & \textbf{0  $\pm$ 0.01 }& \textbf{0 $\pm$ 0.10} & -0.18 $\pm$ 1.74 & -0.02 $\pm$ 0.06 \\
    \hline
    DeepSigns   & -0.01 $\pm$ 0.38 & \textbf{0}  $\pm$ 0.02 & -0.04 $\pm$ 0.12 &  \textbf{0 $\pm$ 0.05}  & \textbf{0} $\pm$ 0.11 & \textbf{0} $\pm$ 0.11 & 0.22 $\pm$ 0.47 & \textbf{0}  $\pm$ 0.05 \\
    \hline
    Uchida   & -2.49 $\pm$ 1.47 & 0.01 $\pm$ 0.28 & -1.96 $\pm$ 0.44 & \textbf{0 }$\pm$ 0.07 & \textbf{0 $\pm$ 0.01} & 0.03 $\pm$ 0.35 & -2.91 $\pm$ 2.56 & \textbf{0} $\pm$ 0.02 \\
    \hline
    DICTION     & \textbf{1.42 $\pm$ 0.82}  & \textbf{0 $\pm$ 0.03}  & \textbf{-1.24 $\pm$ 0.43}  & \textbf{0} $\pm$ 0.06 & \textbf{0} $\pm$ 0.04 & \textbf{0} $\pm$ 0.11 & \textbf{1.70 $\pm$ 1.30}  & \textbf{0 $\pm$ 0.04}  \\
    \hline
    \end{tabular}%
    }
    \label{tab:mean_std_dist}
\end{table}

We also considered the property inference attack (PIA) whose principle is to train a watermark detector $F_{\text{Dect}}$ on several watermarked and non-watermarked neural network models. Again, we assumed the worst-case scenario in which the attacker knows the training data, the exact model architecture of $M^{\text{wat}}$, and the feature extractions key. This threat model is indeed overly strong, but we consider it to demonstrate the effectiveness of our watermark scheme. To perform the property inference attack, we considered the LeNet model we trained over 700 watermarked models and 700 non-watermarked models using MNIST as dataset, all with different keys. We also generated 100 non-watermarked and 100 watermarked models as testing sets, also with different keys. We tested PIA only on LeNet because of the very high complexity of this attack. It required the training of 800 models and the watermarking of each of them. The training set for the detector $F_{\text{Dect}}$ corresponds thus to the watermarking features of each model labeled according to whether or not the model is watermarked. Note that, all generated models were well-trained with good accuracy and watermarks were correctly embedded. As a result, the accuracy of the detector $F_{\text{Dect}}$ does not exceed 60\% on the training and it remains around 50\% for the testing after 50 epochs, which is equivalent to a random guess. This detection accuracy shows that DICTION is resistant to the property inference attack.

\subsection{Integrity \& Efficiency}

The integrity requirement refers to the watermarking scheme's ability to minimize false alarms. In DICTION, this is achieved through the use of a Generative Adversarial Network (GAN) strategy. Specifically, we train the projection model, denoted as $\text{Proj}_{\theta}$, to map the activation maps of the watermarked target model, $M^{\text{wat}}$, to the intended watermark, and those of the original model, $M$, to a random watermark, $b_r$. This approach enables DICTION to effectively meet the integrity requirements.

We also analyzed the overhead incurred by the watermark embedding process in terms of computational and communication complexities. DICTION introduces no communication overhead for watermark embedding, since the embedding process is conducted locally by the model owner. The additional computational cost associated with the embedding of a watermark using DICTION depends on the topology of the target model, $M^{\text{wat}}$, the projection model, $\text{Proj}_{\theta}(\cdot)$, and the number of epochs required to embed the watermark while maintaining high accuracy for the main task. In all of our experiments, we found that 30 epochs were sufficient to allow $\text{Proj}_{\theta}(\cdot)$ to converge with satisfactory accuracy for the target model. As demonstrated in Section \ref{sec:experiments}, the topology of the projection model, $\text{Proj}_{\theta}(\cdot)$, is modest compared to that of the target model, enabling efficient watermark embedding.

Regarding the complexity of watermark extraction, it involves first computing the activation maps, $f^l(\text{LS}, w^{\text{wat}})$, of the images derived from the trigger set, followed by feeding these activation maps into the projection model to extract the watermark. This operation is straightforward, as it does not require a training process to generate the watermark. In Table \ref{tab:time_cost}, we provide the computation time for both training the model and the watermarking process. As shown, DICTION incurs a reasonable overhead, with a runtime overhead of at most 30 epochs for all benchmarks, suggesting high efficiency. However, our embedding complexity remains higher than the methods proposed by Uchida \cite{uchida2017embedding} and DeepSigns \cite{rouhani2019deepsigns} because their projection function relies solely on a multiplication with the secret matrix. On the other hand, our method has similar complexity to ResEncrypt \cite{li2022encryption}, since both methods consist of training an extra model ($\text{Proj}_{\theta}(\cdot)$ and MappingNet; see Section \ref{subsec: whiteWat}), which is less expensive than RIGA \cite{wang2019riga}, as we do not require a third model to preserve the distribution of model weights.

\begin{table}
\centering
\caption{Computation cost for the training and watermarking of all benchmarks}
\label{tab:time_cost}
\begin{tabular}{|c|c|c|c|c|}
\hline
 \multirow{2}{*}{BMs} & \multicolumn{2}{c|}{Training} & \multicolumn{2}{c|}{Watermarking} \\
\cline{2-5}
& Run Time & Epochs & Run Time & Epochs \\
\hline
BM1 & 02:44 min & 50 & 02:00 min & 30 \\
\hline
BM2 & 09:42 min & 100 & 03:09 min & 30 \\
\hline
BM3 & 48:57 min & 200 & 05:39 min & 30 \\
\hline
BM4 & 03:17 min & 50 & 02:22 min & 30 \\
\hline
\end{tabular}
\end{table}

\begin{table*}
    \centering
    \caption{Bit Error Rate (BER) results after watermarking the Benchmark 2 model using different CNN layers and a deep projection function model under various attacks. The table compares the BER after fine-tuning (50 epochs), overwriting (same layers), and pruning 80\% of the model's parameters.  }
    \begin{tabular}{|c|c|c|c|}
        \hline
        CNN Layers & Fine Tuning (50 epochs) & Overwriting (same layers) & Pruning 80\% \\
        \hline
        Fc1 & 0 & 0 & 0 \\
        \hline
        Conv4 & 0 & 0 & 0 \\
        \hline
        Conv3 & 0 & 0 & 0 \\
        \hline
        Mix of all layers & 0 & 0 & 0 \\
        \hline
               Fc1 with a deep Proj$_{\theta}$ & 0 & 0 & 0 \\
        \hline
    \end{tabular}
    \label{tab:res_extension}
\end{table*}

\section{Discussion \& Limitations of the Study}

In this section, we discuss the properties, strengths, and weaknesses of DICTION in general.

\begin{itemize}
    \item \textbf{DICTION paramaterization}: the DICTION scheme depends on several parameters, in particular the layer selected for calculating activation maps and the architecture of the $\text{Proj}_{\theta}$ projection function. In Tables \ref{tab:res_extension}, we have varied all these parameters in the case of BM2, which contains convolutional and fully connected layers in our tests. The results show that regardless of the layer we watermark or the architecture of the $\text{Proj}_{\theta}$ model, our watermark resists all the attacks mentioned above. 
    
    \item \textbf{Trigger Set Generation}: Trigger sets are mainly used for black-box watermarking, where the verifier does not have access to the model's parameters for verification. Various methods have been proposed for generating a trigger set in the state of the art of black-box watermarking methods, especially for out-of-distribution scenarios; working for example with images chosen at random from a public database or images to which a pattern, like a logo, has been added. As in \cite{wen2023tree} for image watermarking generated by a diffusion model, where the images of the trigger set first consist of noise, like that used in DICTION, and then a pattern is added in the Fourier transform of the noise. Another approach involves generating the trigger set with cryptographic primitives, as in \cite{zhu2020secure}, where an image and its label are generated randomly; these images then serve as the seed for a generator of random images and labels. And the images and their associated labels constitute the trigger set.

    \item \textbf{Watermark Generation}: Generally, it is either a random watermark or generated from a  message (e.g., the model owner's identifier). Since sometimes the identifier's encoding space does not match the watermark space, being for instance too long, the hash of the latter plays the role of the watermark.  Another interesting method is to include the original model in the hash calculation to have a watermark that attests both the user's identifier and the link between the original and watermarked models.

    \item \textbf{Activation Maps Calculation}: One can concatenate the activation maps of several layers to create a tensor containing all the information from the selected layers and provide it to $\text{Proj}_\theta$. This strategy allows capturing the information encoded in all the model's layers, but on the downside, it increases the computational complexity of $\text{Proj}_\theta$.  In Tables \ref{tab:res_extension}, we provide the result of DICTION using à $\text{Proj}_\theta$ with three hidden layers and Relu as activation function. The BER is equal to zero agaisnt all the attacks, showing the rebusteness of DICTION.

\item \textbf{Study Limitations}: The watermarking process in DICTION requires training two models, unlike methods like Uchida and DeepSigns, which only require training a single model to perform the watermarking. This is the price for ensuring more robust watermarking requirements. This impact is especially significant for large models like GPT-3, but on the flip side, the watermarking is only done in offline mode. Once the model is watermarked and deployed for users, the watermarking does not disrupt the user experience in terms of computational time or communication complexity. In future work, we would like to test DICTION on large language processing models with real-world applications.

\end{itemize}

\section{Conclusion}
\label{sec:conclusion}
 In this paper, we presented a unified framework that includes all existing white-box watermarking algorithms for DNN models. This framework outlines a theoretical connection between previous works on white-box watermarking for DNN models. From this framework, we derive DICTION, a new white-box Dynamic Robust watermarking scheme that relies on a GAN strategy. Its main originality stands on a watermark extraction function that is a DNN trained using a latent space as an adversarial network. We subjected DICTION to all actual watermark detection and removal attacks and demonstrated that it is much more robust than existing works with lower embedding and extraction complexity. In future work, we would like to broaden the scope of our testing by incorporating additional machine learning applications, such as natural language processing, and learning strategies, such as federated learning.

%\bibliographystyle{IEEEtran}
% argument is your BibTeX string definitions and bibliography database(s)
%\bibliographyd
%\bibliographystyle{IEEEtran}
% \bibliography{references}

\bibliographystyle{unsrt}  
\bibliography{diction}

\end{document}